\documentclass[prd,showpacs,amsmath,showkeys,twocolumn,floatfix,amssymb, preprintnumbers, nofootinbib, superscriptaddress]{revtex4} 
\usepackage{epsfig,dcolumn}
\usepackage{graphicx}
\usepackage{comment} 
\usepackage[usenames]{color}
\usepackage{graphicx}
\usepackage{bm}
 \usepackage{ifpdf}

\def\lsim{\mathrel{\rlap{\lower4pt\hbox{\hskip1pt$\sim$}}
    \raise1pt\hbox{$<$}}}
\def\gsim{\mathrel{\rlap{\lower4pt\hbox{\hskip1pt$\sim$}}
    \raise1pt\hbox{$>$}}}
\begin{document}

\title{  An numerical approach for  finite volume three-body interaction   }

\author{Peng~Guo}
\email{pguo@jlab.org}

\affiliation{Department of Physics and Engineering,  California State University, Bakersfield, CA 93311, USA.}

\author{Vladimir~Gasparian}
\affiliation{Department of Physics and Engineering,  California State University, Bakersfield, CA 93311, USA.}

\date{\today}

\begin{abstract} 

In present work,  we study an numerical approach to one   dimensional finite volume three-body interaction, the method is demonstrated by considering a toy model of  three   spinless   particles interacting with pair-wise   $\delta$-function potentials.  The numerical results are compared with the exact solutions of three spinless bosons interaction when strength of short-range interactions are set equal for all pairs.

  \end{abstract} 

\pacs{ }

\maketitle

\section{Introduction}

Three-particle interaction   plays an important role  in many aspects of hadron/nuclear, atomic and condense matter physics.  The understanding of three-body dynamics is an essential and key element of many physical processes, such as, the decay of  \mbox{$\eta \rightarrow 3 \pi$}   \cite{Kambor:1995yc,Anisovich:1996tx,Colangelo:2009db,Lanz:2013ku,Schneider:2010hs,Kampf:2011wr,Guo:2015zqa,Guo:2016wsi}.     Three-body dynamics in free space has been well studied in the past, many  approaches were developed, for instance,  relativistic Bethe-Salpeter equations approach \cite{Taylor:1966zza,Basdevant:1966zzb,Gross:1982ny},  Faddeev equations method \cite{Faddeev:1960su,Faddeev:1965,Phillips:1966zza,Gloeckle:1983,Fedorov:1993,Gloeckle:1995jg}, and  Khuri-Treiman equation approach \cite{Khuri:1960zz,Bronzan:1963xn,Aitchison:1965kt,Aitchison:1965zz,Aitchison:1966kt,Pasquier:1968zz,Pasquier:1969dt,Guo:2014vya,Guo:2014mpp,Danilkin:2014cra,Guo:2015kla}. However, due to the complication of three-body dynamics in general, the finite volume three-body formalism  is still at its early developing phase.  Recent advances in lattice computation has made the study of hadron scattering possible \cite{Aoki:2007rd,Sasaki:2008sv,Feng:2010es,Dudek:2010ew,Beane:2011sc,Lang:2011mn,Aoki:2011yj,Dudek:2012gj,Dudek:2012xn,Wilson:2014cna,Wilson:2015dqa,Dudek:2016cru}. Unfortunately, because of lacking reliable formalism of three-body interaction in finite volume, the current studies of hadron scattering in lattice QCD are only confined in two-body elastic or inelastic regions. The two-body scattering amplitudes are extracted from the results of  lattice QCD calculations  by using   L\"uscher's formula \cite{Luscher:1990ux} or its extensions to moving frames and to inelastic channels   \cite{Rummukainen:1995vs,Lin:2001ek,Christ:2005gi,Bernard:2007cm,Bernard:2008ax,He:2005ey,Lage:2009zv,Doring:2011vk,Aoki:2011gt,Briceno:2012yi,Hansen:2012tf,Guo:2012hv,Guo:2013vsa}.  An reliable and sensible finite volume three-body formalism  is clearly urgently needed  in lattice QCD community  when the energy levels go above  three-body threshold.

In addition to its application in nuclear/hadron physics, the study of three or more particles  either in free space    or interacting with periodic potentials also has wide applications and interests in condense matter physics.  For an example,  the rapid development of semiconductor   technology has allowed  to manufacture quantum dots (QDs) nanometer sized islands,  and   these new nano-structure materials have triggered a great interest from both experimental and theoretical points of view \cite{Rei02}. Especially, electrons inside these nano-structures can be controlled experimentally, so it may potentially be applied to the development of materials in  quantum computing \cite{Jef02} and spintronics \cite{Gol08}. 
Moreover, the QDs are considered as ideal nano-laboratories to study the physical properties of   few-particle system in  reduced dimensional space.  In this regard, two-electron system \cite{Sch13,Fros13,Sar10,Wang11,Ahn14,Men11,Khr13,Ful11,Fros14,Yak15,Gak13} becomes the simplest arrangement of few-particle systems,  which may be served as a starting point to evaluate the correlation effects on the energy band structure of more complicated systems. Some simplest and exactly solvable models of one quantum dot with two interacting electrons have been studied in the past, {\it e.g.} \cite{Lev92}, in which  the hybridization effects with the states on the leads are also considered.  These early studies show that the effect of the electron-electron interaction, in contrast to the case of non-interacting electrons, indeed changes the electron density of states at the Fermi level, and results in non-trivial corrections to the conductivity and   the negative magnetoresistance in disordered conductors \cite{Alt85}.  Another class of systems, to which the three and more electrons interaction applies and which is of special interest in condense matter physics,  is a many body localization phenomena. In these phenomena, the many-body eigenstates of the Hamiltonian are localized,  and    Anderson localization type of  behavior   can only be described by  the interacting few-particle  dynamics (see, {\it e.g.} \cite{Hus14}).

Many attempts  on finite volume three-body interactions have been made in recent years \cite{Kreuzer:2008bi,Kreuzer:2009jp,Kreuzer:2012sr,Polejaeva:2012ut,Briceno:2012rv,Hansen:2014eka,Hansen:2015zga,Hansen:2016fzj,Hammer:2017uqm,Hammer:2017kms,Guo:2016fgl,Guo:2017ism} from different approaches.  For instance,  quantum field theory based diagrammatic approaches or Faddeev equations  based method  \cite{Kreuzer:2008bi,Kreuzer:2009jp,Kreuzer:2012sr,Polejaeva:2012ut,Briceno:2012rv,Hansen:2014eka,Hansen:2015zga,Hansen:2016fzj}, and  the approach by considering the asymptotic form of wave function in configuration space \cite{Guo:2016fgl,Guo:2017ism}.   Unfortunately, majority of these developments are still mathematically unfriendly to common users and are not easily tested in practice because of complication of three-body dynamics. Only a few limited cases of three-body problem can be solved analytically in low dimension, such as  McGuire's model in finite volume \cite{Guo:2016fgl}. However, diffraction effects in McGuire's model are all cancelled out \cite{McGuire:1964zt},  thus   no new momenta are created over scattering process, though momenta are allowed to be rearranged among three particles.  As the consequence, asymptotic form of wave function contains only plane waves,  the spherical waves  are completely absent due to the cancellation of diffraction effect \cite{Guo:2017ism}. The absence of spherical wave simplifies the algebra of finite volume three-body dynamics dramatically, make it possible to finally have the quantization conditions expressed in  quite a simple way analytically \cite{Guo:2016fgl}. In general cases, the analytic solutions of three-body dynamics are usually not available, even for some seems like simple cases, such as the pair-wise $\delta$-function potentials with unequal strength among pairs \cite{McGuire:1988}. Although, as suggested in \cite{Guo:2017ism}, given the asymptotic form of wave function, it may be  possible to obtain  three-body quantization conditions in an analytic form that involves only on-shell scattering amplitudes,   obtaining analytic asymptotic form  of three-body wave function or parametrization of on-shell three-body scattering amplitudes   never is an easy task even for  ``simple cases'', such as unequal strength $\delta$-function pair-wise interactions. Therefore, in present work, we aim to obtain an numerical approach to finite volume three-body problem. Although the explicit and analytic form of quantization conditions are  sacrificed and abandoned this way,   finite volume three-body problems can be solved numerically and quite reliably  without any approximation, and most importantly, the approach is applicable to general cases even when the three-body forces are included.  To demonstrate the approach, in this work, we consider a simple toy model of three spinless particles interacting with pair-wise $\delta$-function potentials. The exact solutions in finite volume are available   at the limit of equal strength  $\delta$-function potentials \cite{Guo:2016fgl}, which can be used to test our numerical approach. As will be made clear later on, the three-body dynamics is completely determined by Faddeev equations, and  wave functions in both free space and finite box can be constructed from  the solutions of Faddeev equations. The role of  matching condition of   free space and finite volume wave functions is to impose the extra constraints on allowed energy spectra in a   finite box and eventually  leads to discrete values of energy spectra as the consequence of periodic lattice structure. In this work, instead of aiming to obtain   analytic expressions of three-body quantization conditions which may be derived from the matching condition, we propose to search allowed energy spectra numerically by using matching condition  directly. Since three-body dynamics is solely determined by Faddeev equations, and is independent of lattice structure of finite box, the Faddeev equations can thus be solved separately by numerical approach, and solutions may be tabulated and stored   regardless the scattering of particles in free space or finite volume. Then, the  solutions of Faddeev equations may be used as input into matching condition of finite volume problem  to  search for  allowed discrete energy spectra in a finite box.
The strategy of numerical approach is illustrated and tested by a toy model  with particles interacting by pair-wise $\delta$-function potentials, the toy model is solved numerically and the results are compared with the exact solutions at the limit of  equal strength   $\delta$-function potentials among all pairs. At last, we would also like to point out that though our discussion and presentation for finite volume three-body problem has been  focused on pair-wise short-range interactions, the approach can be applied to three-body problems in general when three-body force is also included, and the strategy of solving finite volume three-body problem in general cases remains same.  A brief discussion of three-body problem by including three-body force is presented in Appendix  \ref{Faddeev3bforce}.  For completeness,  a short   review of Faddeev's approach for pair-wise short-range interaction   is also provided in Appendix \ref{Faddeevpair}.

The paper is organized as follows.   In Section \ref{3bfreefinite} we summarize the   formalism of   three-particle interaction in finite volume.   The  numerical approach and results are presented in Section \ref{Numerical}. The summary and discussion are given in Section \ref{summary}.

\section{Three-body interaction for short range interaction  }\label{3bfreefinite}

\subsection{Three-body interaction  in free space}\label{3bfree}

In this work, for the purpose of demonstration of numerical approach,  we consider a non-relativistic  toy model of three-body  interaction in one spatial dimension, by assuming    all  particles are spinless and have equal mass. These assumptions are not essential for physics that we are interested in but only to simplify the algebra and presentation. The interactions among particles are assumed pair-wise and described by   $\delta$-function potentials with strength,     \mbox{$ V_{\alpha \beta}$}, between $\alpha$-th and $\beta$-th particles.   The  three-particle  wave function  satisfies Schr\"odinger  equation,
\begin{align}
& \left [- \frac{1}{2 m} \sum_{i=1}^{3} \frac{d^{2}}{d x_{i}^{2}}  + V_{12}\delta(r_{12}) + V_{23} \delta(r_{23}) + V_{31}\delta(r_{31})  - E \right ]  \nonumber \\
&\quad \quad  \times \Psi(x_{1}, x_{2}, x_{3}; p_{1},p_{2},p_{3})=0.
\end{align}
The three-body problem with pair-wise interactions in free space can be handled by well-known Faddeev's approach    \cite{Faddeev:1960su,Faddeev:1965}.  In this way, the scattering with either free-three-particle or two-body bound state plus third particle in both  initial and final states is treated in the same framework. The details of complete derivations of Faddeev's approach for both scattering of free-three-particle and scattering on a bound state are   listed in   Appendix \ref{Faddeevpair} for the completeness of presentation. A brief discussion for three-body problems with three-body forces is also provided in Appendix    \ref{Faddeev3bforce}.   Hence, only some key results and equations are presented in this section.   As proposed in  \cite{Faddeev:1960su,Faddeev:1965}, the three-body wave function has the form of   \mbox{$\Psi=\Psi_{(0)} +  \sum_{\gamma=1}^{3} \Psi_{(\gamma)}$} if  initial state is free-three-particle, and   \mbox{$\Psi=   \sum_{\gamma=1}^{3} \Psi_{(\gamma)}$} for scattering of third particle on a two-body bound state. The relative  wave functions, $ e^{ -i P R}   \Psi_{(\gamma)} =\psi_{(\gamma)}  (r_{\alpha \beta}, r_{\gamma};q_{ij}, q_{k})  $, are determined by
 \begin{align}
 \psi_{(\gamma)} & (r_{\alpha \beta}, r_{\gamma};q_{ij}, q_{k}) =    \psi^{(in)}_{(\gamma)}  (r_{\alpha \beta}, r_{\gamma};q_{ij}, q_{k})  \nonumber \\
&   + \int_{-\infty}^{\infty}   \frac{d k}{2\pi}   e^{i \sqrt{\sigma^{2}    -\frac{3}{4} k^{2}} | r_{\alpha \beta} | }e^{i k  r_{\gamma}  }    \nonumber \\
&     \quad      \times      i t_{\alpha \beta}(\sqrt{\sigma^{2}   -\frac{3}{4} k^{2} } ) g_{(\gamma)} (k ; q_{ij}, q_{k})    , \nonumber \\
& \quad \quad \quad \quad \quad \quad \quad \quad \quad \quad \quad \quad \alpha \neq \beta \neq \gamma,  \label{freewaveboth}
\end{align}
where  \mbox{$t_{\alpha \beta}(k) =- \frac{ m V_{\alpha \beta}}{2k+i m V_{\alpha \beta}} $}  is the two-body scattering amplitude in pair $(\alpha \beta)$, and \mbox{$\sigma^{2}= m E - \frac{P^{2}}{6}=q_{ij}^{2} + \frac{3}{4} q^{2}_{k}$}.  $ \psi^{(in)}_{(\gamma)} $ is associated to the incoming wave, if initial state is free-three-particle state, it is given by
 \begin{align}
 \psi^{(in)}_{(\gamma)} & (r_{\alpha \beta}, r_{\gamma};q_{ij}, q_{k})     \nonumber \\
&   =  \int_{-\infty}^{\infty}   \frac{d k}{2\pi}   e^{i \sqrt{\sigma^{2}    -\frac{3}{4} k^{2}} | r_{\alpha \beta} | }e^{i k  r_{\gamma}  }      i t_{\alpha \beta}(\sqrt{\sigma^{2}   -\frac{3}{4} k^{2} } )   \nonumber \\
&     \quad    \times  \int_{-\infty}^{\infty}   d r'_{\gamma}    e^{ - i k r'_{\gamma} } \psi_{(0)}(0, r'_{\gamma};q_{ij}, q_{k})         . \label{psiin}
\end{align}
 If the initial state is   incident of i-th particle on a bound state of pair $(jk)$, it is thus given by
   \begin{align}
 \psi^{(in)}_{(\gamma)} & (r_{\alpha \beta}, r_{\gamma};q_{ij}, q_{k}) = \delta_{\gamma, i} \phi^{B}_{(\gamma)}(r_{\alpha \beta}) e^{i q_{i}^{B} r_{\gamma}}       , 
\end{align}
where \mbox{$\phi^{B}_{(\gamma)}(r_{\alpha \beta})  = \sqrt{ - \frac{m V_{j k}}{2}} e^{ \frac{m V_{j k}}{2} |r_{\alpha \beta}| }$} refers to the two-body bound state wave function in pair $( j k)$,  and \mbox{$q_{i}^{B} = \sqrt{ \sigma^{2} + \frac{3}{4} (\frac{m V_{jk}}{2})^{2} }$}.  In either case, the $g_{(\gamma)}$ amplitudes satisfy Faddeev type integral equations.  
In a matrix form, the integral equations for $g_{(\gamma)}$ amplitudes are given by
 \begin{align}
&  G(k  )   
=  G^{(0)}(k )     \nonumber \\
&+ i \int_{-\infty}^{\infty}   \frac{d q}{2\pi}    \frac{2    \sqrt{\sigma^{2}    -\frac{3}{4} q^{2}}      }{ \sigma^{2}    -\frac{3}{4} q^{2} - \left(k + \frac{q}{2} \right )^{2}  + i \epsilon }     \mathcal{K}(\sqrt{\sigma^{2}   -\frac{3}{4} q^{2} } )   G(q  )   ,\label{gampboth}
 \end{align}
 where $G$ and  $G^{(0)}$ stand for column vectors $(  g_{(3)} ,  g_{(1)}, g_{(2)} )^{T}$ and $( g^{(0)}_{(3)} ,  g^{(0)}_{(1)}, g^{(0)}_{(2)} )^{T}$ respectively.  The   matrix $ \mathcal{K}$ is given by Eq.(\ref{gkern}),  
  \begin{align}
&  \mathcal{K}( q) 
=    \begin{bmatrix}
0 &  i t_{  23 } (q) & i t_{ 31} (q)  \\
 it_{12}(q) & 0 & it_{31} (q)    \\
it_{12} (q) & it_{23} (q) & 0
\end{bmatrix}    ,   \nonumber 
 \end{align}
  and   $g^{(0)}_{(\gamma)}$'s  are defined by incoming waves, for an incoming wave of free-three-particle,  we have
\begin{align}
& g^{(0)}_{(\gamma)} (k ;   q_{ij}, q_{k})      =i  \int_{-\infty}^{\infty}   \frac{d q}{2\pi}   \frac{2    \sqrt{\sigma^{2}    -\frac{3}{4} q^{2}}      }{ \sigma^{2}    -\frac{3}{4} q^{2}   - \left(k + \frac{q}{2} \right )^{2}  + i \epsilon }      \nonumber \\
&         \times \left [   i t_{ \beta \gamma}(\sqrt{\sigma^{2}   -\frac{3}{4} q^{2} } )  \int_{-\infty}^{\infty}   d r'_{\alpha}    e^{ - i q r'_{\alpha} } \psi_{(0)}(0, r'_{\alpha};q_{ij}, q_{k}) \right.   \nonumber \\
& \quad \left. + i t_{\gamma \alpha}(\sqrt{\sigma^{2}   -\frac{3}{4} q^{2} } )  \int_{-\infty}^{\infty}   d r'_{\beta}    e^{ - i q r'_{\beta} } \psi_{(0)}(0, r'_{\beta};q_{ij}, q_{k})   \right ]    ,
\end{align}
  and for scattering of    i-th particle by a bound state in pair $(jk)$, thus,
 \begin{align}
 g^{(0)}_{(\gamma)} (k ;   q_{ij}, q_{k})    &   =  \delta_{\alpha, i}  \int d r_{\alpha}  \phi^{B}_{(\alpha)}( r_{\alpha}) e^{ - i  ( k+\frac{ q_{i}^{B} }{2} ) r_{\alpha}}   \nonumber \\
 & +      \delta_{\beta, i}  \int d r_{\beta}  \phi^{B}_{(\beta)}( r_{\beta}) e^{ - i  ( k+\frac{ q_{i}^{B} }{2} ) r_{\beta}}    .
\end{align}

Faddeev type equations, Eq.(\ref{gampboth}), have no analytic solutions due to diffraction effects in general, except the special case when  the strengths of $\delta$-function potential among all pairs are identical: \mbox{$V_{12}=V_{23}=V_{31}=V_{0}$}, see  \cite{Guo:2016fgl}. Nevertheless,  Eq.(\ref{gampboth}) can be solved numerically rather straightforwardly  in general cases, and the numerical solutions of $g_{(\gamma)}$ amplitudes can thus be used as input to construct the   free space three-body wave function by Eq.(\ref{freewaveboth}).  As will be presented next, similarly the finite volume three-body wave function is also constructed by using the solutions of  $g_{(\gamma)}$ amplitudes, see Eq.(\ref{finitewaveboth}).

\subsection{Three-body scattering in finite volume}\label{3bbox}
When particles are confined in a one dimensional periodic box of the size of $L$,  as shown in \cite{Guo:2016fgl}, the relative finite volume wave function must   satisfy periodic boundary condition,
\begin{align}
& \psi^{(L)}(r_{\alpha\beta} + n_{\alpha\beta} L,r_{\gamma} +\frac{1}{2} n_{\alpha\beta} L + n_{\beta\gamma} L ; q_{ij}, q_{k})  \nonumber \\
& \quad     =  e^{ - i \frac{P}{3} n_{\alpha\beta} L} e^{ - i \frac{2 P}{3} n_{\beta \gamma} L}  \psi^{(L)}(r_{\alpha\beta}  ,r_{\gamma} ; q_{ij}, q_{k} ) , \nonumber \\
& \quad \quad \quad \quad \quad \quad   P= \frac{2\pi}{L} d , \ \ \ \   (n_{\alpha\beta}, n_{\beta \gamma}, d) \in \mathbb{Z}  . \label{finitewavefunc}
\end{align}
The finite volume three-body wave function, $\psi^{(L)}$, can be constructed from three-body free space wave function, $\psi$, by
\begin{align}
&  \psi^{(L)} (r_{\alpha\beta}, r_{\gamma}; q_{ij}, q_{k}) =  \sum_{n_{\alpha\beta}, n_{\beta \gamma} \in \mathbb{Z}} e^{i \frac{P}{3} n_{\alpha\beta} L } e^{i \frac{2P}{3} n_{\beta \gamma} L }  \nonumber \\
& \quad \quad \times  \psi(r_{\alpha\beta} + n_{\alpha\beta} L, r_{k}  +\frac{1}{2} n_{\alpha\beta} L + n_{\beta \gamma} L; q_{ij}, q_{k}).  \label{relwaveconstr}
\end{align} 
The infinite sum may be carried out  by using relation
\begin{align}
&  \sum_{n_{\alpha \beta} \in \mathbb{Z}} e^{i ( \frac{P}{3}  + \frac{k}{2} ) n_{\alpha \beta} L }    e^{i \sqrt{\sigma^{2}    -\frac{3}{4} k^{2}} | r_{\alpha \beta} + n_{\alpha \beta} L | }    \nonumber \\
&=e^{i \sqrt{\sigma^{2}    -\frac{3}{4} k^{2}} | r_{\alpha \beta}  | } + \frac{e^{i \sqrt{\sigma^{2}    -\frac{3}{4} k^{2}}  r_{\alpha \beta}   }  }{e^{- i ( \sqrt{\sigma^{2}    -\frac{3}{4} k^{2}}  +\frac{P}{3}  + \frac{k}{2}  ) L} -1}   \nonumber \\
&  +  \frac{e^{ - i \sqrt{\sigma^{2}    -\frac{3}{4} k^{2}}  r_{\alpha \beta}   }  }{e^{- i ( \sqrt{\sigma^{2}    -\frac{3}{4} k^{2}}  - \frac{P}{3}  - \frac{k}{2}  ) L} -1}.
\end{align}
Also using the Poisson summation formula, \mbox{$\sum_{  n_{\beta \gamma} \in \mathbb{Z}}  e^{i  ( \frac{2P}{3} + k) n_{\beta \gamma} L }        = \frac{2\pi}{L} \sum_{n \in \mathbb{Z}} \delta ( \frac{2P}{3} + k - \frac{2\pi }{L} n )$} and free space three-body wave function in Eq.(\ref{freewaveboth}),   the finite volume three-body wave function hence yields
  \begin{align}
&  \psi^{(L)}_{(\gamma)}   (r_{\alpha \beta}, r_{\gamma};q_{ij}, q_{k})  
  =    \frac{1}{L} \sum_{n \in \mathbb{Z}}^{k=-\frac{2P}{3}+ \frac{2\pi}{L} n}   \left [   e^{i \sqrt{\sigma^{2}    -\frac{3}{4} k^{2}} | r_{\alpha \beta}   | }         e^{i k r_{\gamma}   }    \right. \nonumber \\
 &   \left.  + \frac{e^{i \sqrt{\sigma^{2}    -\frac{3}{4} k^{2}}  r_{\alpha \beta}   }   e^{i k r_{\gamma}   }}{e^{- i ( \sqrt{\sigma^{2}    -\frac{3}{4} k^{2}}  +\frac{P}{3}  + \frac{k}{2}  ) L} -1}    +  \frac{e^{ - i \sqrt{\sigma^{2}    -\frac{3}{4} k^{2}}  r_{\alpha \beta}   }   e^{i k r_{\gamma}   }}{e^{- i ( \sqrt{\sigma^{2}    -\frac{3}{4} k^{2}}  - \frac{P}{3}  - \frac{k}{2}  ) L} -1} \right ]  \nonumber \\
&  \quad    \times  i t_{\alpha \beta}(\sqrt{\sigma^{2}   -\frac{3}{4} k^{2} } )   g_{(\gamma)} (k ; q_{ij}, q_{k}) ,\nonumber \\
& \quad \quad \quad \quad \quad \quad \quad \quad \quad \quad \quad \quad\quad \quad \quad \quad \alpha \neq \beta \neq \gamma   . \label{finitewaveboth}
\end{align}

The quantization conditions that yield the discrete energy spectra for three-body interaction in a finite box may be obtained by matching condition \cite{Guo:2016fgl}, 
\begin{align}
& \sum_{\gamma=1}^{3} \left [  \psi^{(L)}_{(\gamma)}  (r_{\alpha \beta}, r_{\gamma};q_{ij}, q_{k}) -  \psi_{(\gamma)}  (r_{\alpha \beta}, r_{\gamma};q_{ij}, q_{k})  \right ]  \nonumber \\
& =  \begin{cases}
     \psi_{(0)}  (r_{\alpha \beta}, r_{\gamma};q_{ij}, q_{k}),       &   \text{if free-particle initial state,}  \\
    0, &   \text{if incident on a bound state} , \\
  \end{cases} \label{matchcondition}
\end{align}
where both finite volume wave function, $\psi^{(L)}_{(\gamma)} $ in Eq.(\ref{finitewaveboth}), and free space wave function, $\psi_{(\gamma)} $ in Eq.(\ref{freewaveboth}), are determined by the solutions of Faddeev equations in Eq.(\ref{gampboth}). In another word, the three-body dynamics is completely described by Faddeev equations, and the role of quantization conditions or matching conditions in Eq.(\ref{matchcondition}) is nothing but to impose constraints on allowed       energy spectra to meet the requirement of periodic lattice structure in a finite box. Therefore, given the solution of Faddeev equations of $g$-amplitude, the task is to scan all the possible combination of $(q_{ij}, q_{k})$ to find the solution of energy spectrum that fulfill the matching condition in Eq.(\ref{matchcondition})  for an arbitrary pair of $(r_{\alpha\beta}, r_{\gamma})$. Bearing this in mind, instead of finding the basis of  asymptotic form of wave function in both free space and finite volume \cite{Luscher:1990ux} and deriving an analytic expression of secular equation from matching conditions \cite{Guo:2016fgl,Guo:2017ism}, our strategy is to solve Faddeev equations first, and use the solutions of $g_{(\gamma)}$-amplitudes  as input of matching condition, Eq.(\ref{matchcondition}), to search all possible allowed energy spectra of three-body interaction numerically.  Although, it seems like the analytic forms of secular equation is lost this way, the numerical approach presented in this work is rather straightforward, and the formalism itself is rather simple and user friendly. The only trade-off is that Faddeev equations has to be solved first numerically, and therefore it is more computationally involved. Fortunately,    solving Faddeev equations and searching allowed energy spectra in matching conditions are two independent processes,  they can be carried out separately. Therefore in practice, it may be plausible to solve Faddeev equations first for multiple  initial momenta and energies, and  then proceed with second step of energy spectra searching by using matching condition in Eq.(\ref{matchcondition}). In addition, the procedure and strategy of solving finite volume three-body problem is not   limited to only   pair-wise interactions, but also could be applied to general cases with three-body forces, see discussion in Appendix  \ref{Faddeev3bforce}. The idea is demonstrated and compared to exact solutions in next section.

\section{Numerical test   and exact solutions at the limit of \mbox{$V_{12}=V_{23}=V_{31}=V_{0}$}}\label{Numerical} 

\subsection{Scattering of three-boson in general}\label{3bgen}

Let's consider a totally symmetric  free-three-particle   incoming wave,
\begin{equation}
\psi^{sym}_{(0)} =\sum_{k=1}^{3} \left ( e^{i q_{ij} r_{12}}+e^{-i q_{ij} r_{12}} \right )  e^{ i q_{k} r_{3}}. \label{psi0sym}
\end{equation}
At the limit of \mbox{$V_{12}=V_{23}=V_{31}=V_{0}$},   it may describe the scattering of three spinless bosons.
Hence, we obtain
\begin{align}
& g^{(0)}_{(\gamma)} (k ;   q_{ij}, q_{k})     \nonumber \\
&  =- 2 i    \frac{2   q_{12}    \left [   i t_{ \beta \gamma}( q_{12} )     + i t_{\gamma \alpha}( q_{12} )    \right ]  }{(k- q_{2} - i \epsilon) (k - q_{1} + i \epsilon) }      \nonumber \\
&          + 2 i    \frac{2   q_{23}    \left [   i t_{ \beta \gamma}(- q_{23} )     + i t_{\gamma \alpha}(- q_{23} )    \right ]  }{(k- q_{2} - i \epsilon) (k - q_{3} - i \epsilon) }      \nonumber \\
&          + 2 i    \frac{2   q_{31}    \left [   i t_{ \beta \gamma}(- q_{31} )     + i t_{\gamma \alpha}(- q_{31} )    \right ]  }{(k- q_{3} - i \epsilon) (k - q_{1} + i \epsilon) }      \nonumber \\ 
&          + 4 \pi \delta(k-q_{3})    \left [   i t_{ \beta \gamma}(- q_{23} )     + i t_{\gamma \alpha}(- q_{23} )    \right ]     .  \label{symg0}
\end{align}
 Normally, it is more stable numerically to separate the $\delta$-function type singular terms by redefining $g$'s amplitudes,
\begin{align}
g_{(\gamma)} & (k ;   q_{ij}, q_{k})   = \hat{g}_{(\gamma)} (k ;   q_{ij}, q_{k})  \nonumber \\
&+  4 \pi \delta(k-q_{3})    \left [   i t_{ \beta \gamma}(- q_{23} )     + i t_{\gamma \alpha}(- q_{23} )    \right ] , \label{gghat}
\end{align}
where similar to equations of $g_{(\gamma)} $'s,   integral equations for  $\hat{g}_{(\gamma)} $'s are given by,
\begin{align}
&  \hat{G}(k  )   
=  \hat{G}^{(0)}(k )     \nonumber \\
&+ i \int_{-\infty}^{\infty}   \frac{d q}{2\pi}    \frac{2    \sqrt{\sigma^{2}    -\frac{3}{4} q^{2}}      }{  \sigma^{2}    -\frac{3}{4} q^{2}   - \left(k + \frac{q}{2} \right )^{2}  + i \epsilon }     \mathcal{K}(\sqrt{\sigma^{2}   -\frac{3 }{4}q^{2}  } )   \hat{G}(q  )   ,  \label{Ghat}
 \end{align}
 where $\hat{G}$ and  $\hat{G}^{(0)}$ stand for column vectors $(  \hat{ g}_{(3)} ,  \hat{ g}_{(1)}, \hat{ g}_{(2)} )^{T}$ and $(  \hat{ g}^{(0)}_{(3)} ,  \hat{ g}^{(0)}_{(1)}, \hat{ g}^{(0)}_{(2)} )^{T}$ respectively,  and
\begin{align}
& \hat{g}^{(0)}_{(\gamma)} (k ;   q_{ij}, q_{k})     \nonumber \\
&  =- 2 i    \frac{2   q_{12}       i t_{ \beta \gamma}( q_{12} ) \left [1+ i t_{\gamma \alpha} (-q_{23}) + i t_{\alpha\beta} (-q_{23}) \right ]      }{(k- q_{2} - i \epsilon) (k - q_{1} + i \epsilon) }      \nonumber \\
&  - 2 i    \frac{2   q_{12}      i t_{\gamma \alpha}( q_{12} )   \left [1+ i t_{\alpha\beta} (-q_{23}) + i t_{\beta \gamma  } (-q_{23})  \right ]       }{(k- q_{2} - i \epsilon) (k - q_{1} + i \epsilon) }      \nonumber \\
&          + 2 i    \frac{2   q_{23}    \left [   i t_{ \beta \gamma}(- q_{23} )     + i t_{\gamma \alpha}(- q_{23} )    \right ]  }{(k- q_{2} - i \epsilon) (k - q_{3} - i \epsilon) }      \nonumber \\
&          + 2 i    \frac{2   q_{31}    \left [   i t_{ \beta \gamma}(- q_{31} )     + i t_{\gamma \alpha}(- q_{31} )    \right ]  }{(k- q_{3} - i \epsilon) (k - q_{1} + i \epsilon) }         .  
\end{align}

In terms of $\hat{g}_{(\gamma)} $ amplitudes, the totally symmetric free space and finite volume wave functions are determined  respectively by,
\begin{align}
&  \psi^{sym}_{(\gamma)}  (r_{\alpha \beta}, r_{\gamma};q_{ij}, q_{k})   \nonumber \\
&=   2 i t_{\alpha \beta}(q_{12})  \left [1+   i t_{ \beta \gamma}(- q_{23} )     + i t_{\gamma \alpha}(- q_{23} )    \right ]     e^{i q_{12}  | r_{\alpha \beta} | }e^{i q_{3}  r_{\gamma}  }    \nonumber \\
  &   +  2 i t_{\alpha \beta}(- q_{23})    e^{-i q_{23}  | r_{\alpha \beta} | }e^{i q_{1}  r_{\gamma}  }  \nonumber \\
  & +  2 i t_{\alpha \beta}(- q_{31})    e^{-i q_{31}  | r_{\alpha \beta} | }e^{i q_{2}  r_{\gamma}  }    \nonumber \\
&   + \int_{-\infty}^{\infty}   \frac{d k}{2\pi}   e^{i \sqrt{\sigma^{2}    -\frac{3}{4} k^{2}} | r_{\alpha \beta} | }e^{i k  r_{\gamma}  }    i t_{\alpha \beta}(\sqrt{\sigma^{2}   -\frac{3}{4} k^{2} } )     \nonumber \\
&     \quad   \quad  \times   \hat{g}_{(\gamma)} (k ; q_{ij}, q_{k})    ,  \quad \quad \quad \quad\quad \quad     \alpha \neq \beta \neq \gamma   , \label{freewavetotsym}
\end{align}
and separating the $\delta$-function type singular terms from $g$'s by Eq.(\ref{gghat}) has no effects on non-trivial solutions of three-body problem in finite volume, thus for non-trivial solutions, finite volume wave function has the similar form as in Eq.(\ref{finitewaveboth}),
  \begin{align}
&  \psi^{sym (L)}_{(\gamma)}   (r_{\alpha \beta}, r_{\gamma};q_{ij}, q_{k})  
   \nonumber \\
   &=    \frac{1}{L} \sum_{n \in \mathbb{Z}}^{k=-\frac{2P}{3}+ \frac{2\pi}{L} n}   \left [   e^{i \sqrt{\sigma^{2}    -\frac{3}{4} k^{2}} | r_{\alpha \beta}   | }         e^{i k r_{\gamma}   }    \right. \nonumber \\
 &   \left.  + \frac{e^{i \sqrt{\sigma^{2}    -\frac{3}{4} k^{2}}  r_{\alpha \beta}   }   e^{i k r_{\gamma}   }}{e^{- i ( \sqrt{\sigma^{2}    -\frac{3}{4} k^{2}}  +\frac{P}{3}  + \frac{k}{2}  ) L} -1}    +  \frac{e^{ - i \sqrt{\sigma^{2}    -\frac{3}{4} k^{2}}  r_{\alpha \beta}   }   e^{i k r_{\gamma}   }}{e^{- i ( \sqrt{\sigma^{2}    -\frac{3}{4} k^{2}}  - \frac{P}{3}  - \frac{k}{2}  ) L} -1} \right ]   \nonumber \\
&  \quad    \times   i t_{\alpha \beta}(\sqrt{\sigma^{2}   -\frac{3}{4} k^{2} } )  \hat{g}_{(\gamma)} (k ; q_{ij}, q_{k}) , \nonumber \\
& \quad \quad \quad \quad \quad \quad \quad \quad \quad \quad \quad \quad\quad \quad \quad \quad \alpha \neq \beta \neq \gamma   .  \label{fvwavetotsym}
\end{align}

As discussed in previous section, when strength of potentials are not identical,  Faddeev equations have no analytic solutions, and have to be solved numerically.  Then solutions of  $\hat{g}_{(\gamma)} $ amplitudes by solving Eq.(\ref{Ghat}) equations can be used as input to construct both free space and finite volume wave function according to Eq.(\ref{freewavetotsym}) and Eq.(\ref{fvwavetotsym}). Finally, the discrete spectra of three-body interaction in finite volume may be searched numerically by using matching condition, Eq.(\ref{matchcondition}).

\subsection{Exact solutions for  equal strength $\delta$-function potentials: $V_{12}=V_{23}=V_{31}=V_{0}$}\label{3bexact}
In the case of equal strength of $\delta$-function potentials, \mbox{$V_{12}=V_{23}=V_{31}=V_{0}$}, the three-body interaction in finite volume is exactly solvable \cite{Guo:2016fgl}.  For totally symmetric incoming wave, see Eq.(\ref{psi0sym}),   the exact solutions of $g$-amplitude are
\begin{align}
 g_{( 1,2,3)} & (k;   q_{ij}, q_{k} )  = 8 \pi \delta(k-q_{3})     i t(- q_{23} )      \nonumber \\
&+  \frac{\left (1+ \frac{  \frac{i m V_{0}}{2}}{ \sqrt{\sigma^{2} - \frac{3}{4} k^{2}} } \right )  \frac{  (-2  m V_{0}) 6 k   }{ \left (1+ \frac{ i m V_{0}}{2 q_{12}} \right )   \left (1-  \frac{ i m V_{0}}{2 q_{23}} \right )  \left (1- \frac{ i m V_{0}}{2 q_{31}} \right )  } }{  (k- q_{3} - i \epsilon)(k- q_{2} - i \epsilon) (k - q_{1} + i \epsilon)}, \label{gsymexact}
\end{align}
where \mbox{$ t(q) = - \frac{ m V_{0}}{2 q + i m V_{0}}$} refers to two-body scattering amplitude.

 The totally symmetric wave function is expressed in terms of a single independent coefficient, see \cite{Guo:2016fgl},
 \begin{align}
& \psi^{sym} (r_{12},r_{3}; q_{ij}, q_{k})  \nonumber \\
& =\left ( A^{sym}   (r_{12},r_{3}) e^{i q_{12} r_{12}} + A^{sym}    (-r_{12},r_{3})  e^{ - i q_{12} r_{12}} \right ) e^{i q_{3} r_{3}}  \nonumber \\
&+ \left ( A^{sym}   (r_{31},r_{2}) e^{i q_{23} r_{12}} + A^{sym}  (-r_{23},r_{1}) e^{ - i q_{23} r_{12}} \right ) e^{i q_{1} r_{3}} \nonumber \\
&+ \left ( A^{sym}  (r_{23},r_{1})  e^{i q_{31} r_{12}} + A^{sym}  (-r_{31},r_{2}) e^{ - i q_{31} r_{12}} \right ) e^{i q_{2} r_{3}}, \label{wavesolsym}
\end{align}
where \mbox{$r_{23} =  -\frac{r_{12}}{2} + r_{3} $}  and  \mbox{$r_{31} =  -\frac{r_{12}}{2} - r_{3}$}, and
\begin{align}
A^{sym}   & (r_{12},r_{3}) = 1 + \theta(r_{12})   2  it(q_{12}) \left [ 1+ 2 i t(-q_{23})\right ]   \nonumber \\
&+    \theta(-r_{23})  2  it(-q_{23}) +  \theta(-r_{31})  2  it(-q_{31})      \nonumber \\
  &-     \theta(r_{12})  \theta(r_{23}) 4    i  \mathbf{ T}_{1}      + \theta(r_{12}) \theta(-r_{31})   4   i  \mathbf{ T}_{2}         ,\label{coefwavesolsym}
\end{align}
and
 \begin{align}
 i \mathbf{ T}_{1}   &=\frac{  \left ( \frac{i m V_{0}}{2 q_{23}} \right )    \left ( \frac{ i  m V_{0}}{2 q_{31}}  - \frac{ i  m V_{0}}{2 q_{12}} \right ) }{  \left ( 1+ \frac{ i  m V_{0}}{2 q_{12}} \right ) \left (1- \frac{ i  m V_{0}}{2 q_{23}} \right ) \left (1- \frac{ i  m V_{0}}{2 q_{31}} \right )  }    ,  \nonumber \\
 i \mathbf{ T}_{2}    &=\frac{  \left ( \frac{i m V_{0}}{2 q_{31}} \right ) \left ( \frac{ i  m V_{0}}{2 q_{23}} - \frac{ i  m V_{0}}{2 q_{12}} \right )  }{\left (1+ \frac{i m V_{0}}{2 q_{12}} \right )\left (1- \frac{i m V_{0}}{2 q_{23}} \right )\left (1- \frac{i m V_{0}}{2 q_{31}} \right )}   .
 \end{align}
We remark that the term $\psi_{(0)}^{sym}$ in Eq.(34) presented in \cite{Guo:2016fgl} was a typo, see Eq.(\ref{wavesolsym}) for  the correct version above. The totally symmetric finite volume wave function has the same structure as free space wave function given  in Eq.(\ref{wavesolsym}), the coefficient in finite volume is given by 
\begin{align}
 A^{sym(L)}  (r_{12},r_{3})     
&= 4    i  \mathbf{ T}_{2}   \left [ \theta(r_{12}) +   \frac{ e^{ i (\frac{2}{3} P +q_{2})L}}{1- e^{ i (\frac{2}{3} P +q_{2})L}}       \right ]  \nonumber \\
 & \quad \quad \  \times  \left [  \theta(-r_{31}) +  \frac{ e^{ i (\frac{2}{3} P +q_{3})L}}{1- e^{ i (\frac{2}{3} P +q_{3})L}}  \right ]    \nonumber \\
&       -     4     i  \mathbf{ T}_{1}  \left [ \theta(r_{12}) + \frac{ e^{- i (\frac{2}{3} P +q_{1})L}}{1- e^{- i (\frac{2}{3} P +q_{1})L}}     \right ]   \nonumber \\
& \quad \quad \ \times  \left [     \theta(r_{23})  + \frac{ e^{ i (\frac{2}{3} P +q_{3})L}}{1- e^{ i (\frac{2}{3} P +q_{3})L}} \right ]  . \label{boxcoefwavesolsym}
\end{align}

As discussed in \cite{Guo:2016fgl},    the quantization conditions are obtained by considering the matching condition between free space wave function and finite volume wave function, and for the case of equal strength $\delta$-function potentials, quantization conditions are given in simple forms,
\begin{align}
& \cot (\frac{P}{3} + \frac{q_{3}}{2})L + \cot  \left (  \phi (-q_{31})-\phi (-q_{23})  \right ) =0 , \nonumber \\
&  \cot (\frac{P}{3} + \frac{q_{1}}{2})L + \cot  \left ( - \phi (-q_{31}) - \phi (q_{12}) \right ) =0, \nonumber \\
&  \cot (\frac{P}{3} + \frac{q_{2}}{2})L + \cot  \left ( \phi (-q_{23}) + \phi (q_{12}) \right ) =0 , \label{seculareq}
\end{align}
where two-body phase shift  is given by \mbox{$\phi(q) =\cot^{-1} \left ( - \frac{2 q}{mV_{0}} \right ) $}.

  \begin{figure}
\begin{center}
\includegraphics[width=3.8 in,angle=0]{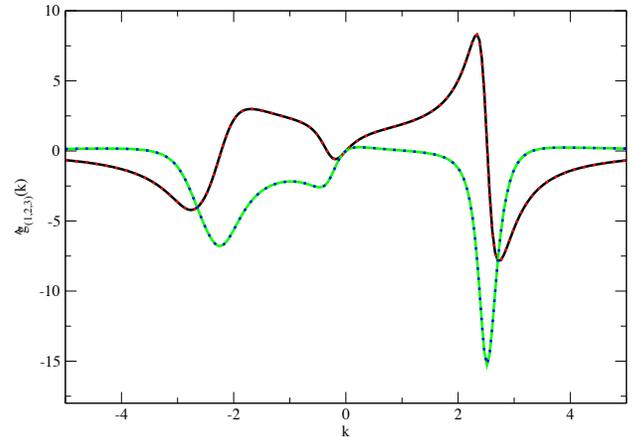}  
\caption{ The comparison of numerical solutions of  integral Eq.(\ref{Ghat}), $\hat{g}_{(1,2,3)}$, with exact solutions given in Eq.(\ref{gsymexact}).  Solid black and solid green curves represent real and imaginary parts of numerical solutions, and dotted red and blue are real and imaginary parts of exact solutions respectively. The parameters of the toy model are chosen as $mV_{0}=2.0$, $q_{12} =1.0+0.4 i$ and $q_{3} =2.5+ 0.2 i$, where  an imaginary part is given to both $(q_{12},q_{3})$ to smooth out the curves near the pole position for a better visualization purpose only.   \label{fig:gsym}  } 
\end{center}
\end{figure}

  \begin{figure}
\begin{center}
\includegraphics[width=3.8 in,angle=0]{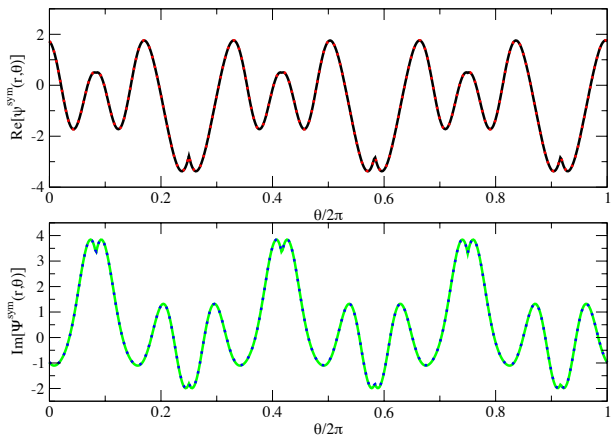}  
\caption{ The comparison of numerical solution of   free space wave function, $\psi^{sym}$,  constructed by using the solution of Faddeev equations and Eq.(\ref{freewavetotsym})   with exact solutions given in Eq.(\ref{wavesolsym}-\ref{coefwavesolsym}).  The real and imaginary parts of free space wave function are presented in upper and lower panels respectively.  Solid black and dotted red curves in upper panel represent real part of numerical solution and exact solution respectively, and solid green and dotted blue curves in lower panel represent imaginary part of numerical solution and exact solution respectively. The parameters are chosen as $mV_{0}=2.0$, $q_{12} =1.0$ and $q_{3} =2.5$. The wave function as function of $(r, \theta)$ is plotted with a fixed $r=5.5$.   \label{fig:wavefree}  } 
\end{center}
\end{figure}

  \begin{figure}
\begin{center}
\includegraphics[width=3.8 in,angle=0]{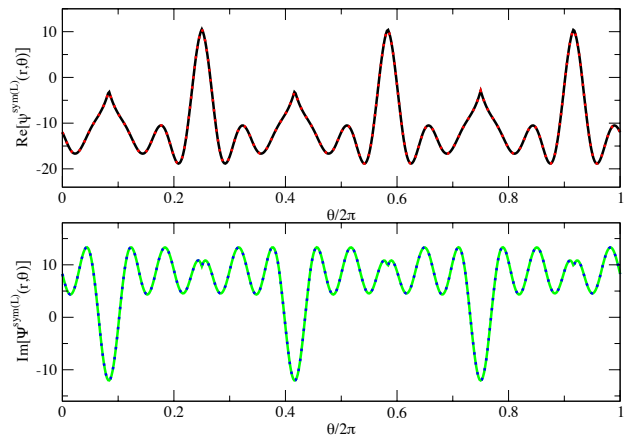}  
\caption{ The comparison of numerical solution of center of mass  \mbox{$(P=0)$} finite volume wave function, $\psi^{sym(L)}$,  constructed by using the solution of Faddeev equations and Eq.(\ref{fvwavetotsym})   with exact solutions given by Eq.(\ref{boxcoefwavesolsym}).  The real and imaginary parts of finite volume wave function are presented in upper and lower panels respectively.  The colors coding, line styles and parameters are the  same as  in Fig.\ref{fig:wavefree}.  The finite volume wave function as function of $(r, \theta)$ is plotted with a fixed $r=5.5$.  
 \label{fig:waveLat}  } 
\end{center}
\end{figure}

\subsection{Strategy of searching allowed energy spectra}

The discrete energy spectra in finite volume are determined by the matching condition of wave functions in free space and finite volume, such as   Eq.(\ref{matchcondition}). Therefore, in principle, the task of obtaining three-body energy spectra in finite volume is thus to search all possible combination of $(q_{ij}, q_{k})$,  so that, the matching condition, Eq.(\ref{matchcondition}), is satisfied for an arbitrary $(r_{12},r_{3})$.

Normally, in order to explicitly removing $(r_{12},r_{3})$ dependence in matching condition, the quantization conditions may be further derived by expanding the wave functions in terms of certain orthogonal basis, see \cite{Guo:2017ism}. For example,  the choice of basis may be made based on the asymptotic behavior of three-body wave functions \cite{Guo:2017ism}, such as  Bessel functions, $\{ J_{J}(\sigma r), N_{J} (\sigma r)\}$, and $e^{i J \theta}$ in $(r_{12},r_{3})$ plane, where $(r,\theta)$ are the radius and polar angle of coordinate, $(r_{12},r_{3})$, respectively.  Therefore, according to asymptotic behaviors of three-body wave function, the wave functions in free space and finite volume normally   have the forms, see \cite{Guo:2017ism},
\begin{align}
& \psi(r_{12},r_{3}; q_{ij},q_{k}) = \sum_{J,J'}  e^{i J \theta} \nonumber \\
& \quad \times  \left [  c_{J, J'} (q_{ij},q_{k}) J_{J'} (\sigma r) + \delta_{J,J'} d_{J}(q_{ij},q_{k})  N_{J'}(\sigma r) \right ],    \\
& \psi^{(L)}(r_{12},r_{3}; q_{ij},q_{k}) = \sum_{J,J'} e^{i J \theta}  \nonumber \\
&\quad \times \left [   c^{(L)}_{J, J'} (q_{ij},q_{k}) J_{J'} (\sigma r)+ \delta_{J,J'} d_{J}(q_{ij},q_{k})  N_{J'}(\sigma r) \right ].
\end{align}
Hence, the matching condition, $\psi=\psi^{(L)}$, leads to the quantization conditions that are given by determinant condition in terms of expansion coefficients alone,
\begin{align}
\det \left [ c_{J, J'} (q_{ij},q_{k})  - c^{(L)}_{J, J'} (q_{ij},q_{k}) \right ] =0. \label{detcond}
\end{align}

Unfortunately, except a few special cases, such as, equal strength $\delta$-function potentials, the quantization conditions  usually do not possess  an simple analytic expression and appear messy and complicated. Moreover, the expansion has to  be truncated in practice to solve determinant condition,  thus, the convergence of expansion somehow more or less depends on the choice of expansion basis of wave functions. Therefore, instead of making  efforts on obtaining quantization conditions and solving determinant condition, such as in Eq.(\ref{detcond}), we propose to solve the matching condition directly.  Though,  the explicit expression of quantization conditions is sacrificed,   the procedure and strategy of obtaining discrete energy spectra is actually more clear and straightforward.  First of all, the Faddeev equations that define the dynamics of three-body interaction, such as in Eq.(\ref{gampboth}), are solved numerically.  Next, the solutions of Faddeev equations are used as input to construct both free space and finite volume wave functions, see Eq.(\ref{freewaveboth}) and Eq.(\ref{finitewaveboth}). At last, the discrete energy spectra that are determined by periodic lattice structure are obtained by searching matching condition, $\psi=\psi^{(L)}$, directly.

  \begin{figure}
\begin{center}
\includegraphics[width=3.8 in,angle=0]{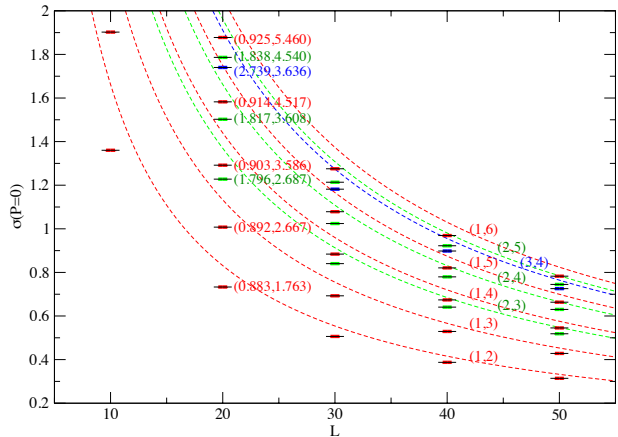}  
\caption{  The center of mass \mbox{$(P=0)$}  three-body energy spectra, \mbox{$\sigma = \sqrt{mE}= \sqrt{ q_{12}^{2} + \frac{3}{4}q_{3}^{2} }$}, as function of lattice size, $L$. The solid black bars and colored boxes represent solutions of secular equation, Eq.(\ref{seculareq}), and matching condition by solving $\mathcal{M}  ( q_{12}, q_{3})=0$  respectively.   The red, green and blue dashed curves represent the non-interacting energy spectra: $ \frac{2\pi}{L} \sqrt{ \left ( \frac{n_{1}-n_{2}}{2} \right )^{2} + \frac{3}{4}\left (  n_{1}+n_{2} \right )^{2} }$ with $(n_{1},n_{2}) \in \mathbb{Z}$.  The colored lines are labeled by pari of integers, $(n_{1},n_{2})$, which are associated to momenta of non-interacting particles by relations: $p_{1,2,3} = \frac{2\pi}{L} n_{1,2,3}$. As examples of both discrete  energy and momenta in finite box, the three-body energy levels for $L=20$ are also labeled by pair of discrete values of $(q_{12},q_{3})$  next to corresponding energy level.   \label{fig:spectra}  } 
\end{center}
\end{figure}

In this subsection, using scattering of three-boson at the limit of equal strength $\delta$-function potentials as an numerical test, we solve Faddeev equations  presented in subsection \ref{3bgen}, and use the solution of $g$-amplitude as input to construct wave functions and further seek the discrete spectra that satisfy  the matching condition in Eq.(\ref{matchcondition}). The numerical results are compared with exact solutions. 
The comparison of  the numerical solutions of $g$-amplitudes, free space wave function and finite volume wave function with  exact solutions are presented in Fig.\ref{fig:gsym}-\ref{fig:waveLat} respectively. The matching condition, $\psi=\psi^{(L)}$, is solved numerically by root finding method, more specifically,  a   function,  $\mathcal{M}  ( q_{12}, q_{3})$,  is introduced,
\begin{align}
 &\mathcal{M}  ( q_{12}, q_{3}) \nonumber \\
   & =  \frac{1}{N} \sum_{(r_{12},r_{3})=1}^{N}  \frac{ \left |  \psi   (r_{12}, r_{3};q_{12}, q_{3})  -   \psi^{(L)}  (r_{12 }, r_{3};q_{12}, q_{3})     \right | }{\left |  \psi   (r_{12}, r_{3};q_{12}, q_{3})  \right |}, 
\end{align}
where sum of $(r_{12},r_{3})$ are carried out by choosing some discrete values,  for example, in present work, $(r_{12},r_{3})$ space is discretized in terms of polar coordinate, $(r, \theta)$. About $30$ points of $r$ values in range $r\in [1,5]$ and $100$ points of $\theta \in [0,2 \pi]$  are  taken in the sum. $N$ refers to the total numbers of discrete points of $(r_{12},r_{3})$ in the sum. For the non-trivial solutions of three-body energy spectra (none of particle momentum coincides with $\frac{2\pi}{L} n, n\in \mathcal{Z}$),  the possible discrete values of pair $(q_{12},q_{3})$ are searched by  performing   root finding of condition,  $\mathcal{M}  ( q_{12}, q_{3})=0$. The results are compared with the   solutions given by quantization condition in Eq.(\ref{seculareq}), and   presented in Fig.\ref{fig:spectra}. We remark that the solutions of quantization condition for \mbox{$q_{ij}=0$} are excluded in Fig.\ref{fig:spectra}, it is not difficult to see that the wave functions vanish due to the symmetry of three-body for \mbox{$q_{ij}=0$}  at the limit of \mbox{$V_{12}=V_{23}=V_{31}=V_{0}$}, thus, the solutions for \mbox{$q_{ij}=0$}  are considered as trivial and not included in Fig.\ref{fig:spectra}.

\section{Discussion and conclusion}\label{summary}

As mentioned in previous sections,  the dynamics of three-body interaction is completely determined by Faddeev equations regardless the   three-particle interacting in free space or finite box.  When the periodic boundary condition is considered,   the allowed energy spectra are constrained by matching condition between free space three-body wave function and finite volume wave function, and eventually become discrete. Therefore, seeking discrete three-body energy spectra in finite box can be carried out by two independent procedures, first of all, solving dynamical Faddeev equations and using the solutions of Faddeev equations to construct wave functions. Secondly, in stead of seeking an analytic expression of quantization conditions for finite volume three-body interaction that may be derived from matching condition of three-body wave functions, we propose to search for discrete energy spectra by using matching condition   directly. 

In this work, we demonstrated this approach by considering scattering of three spinless bosons interacting with $\delta$-function potentials. In this case, the exact analytic solutions exist  at the limit of equal strength of $\delta$-function potential among all pairs. The finite volume three-body interaction in this toy model is then solved numerically, and the discrete energy spectra are searched by using matching condition of wave functions. Finally, all the numerical results are compared with exact solutions presented in section \ref{Numerical}. At last, we want to stress that although a specific toy model with only pair-wise interaction is solved in this work, this approach is in fact not limited to only pair-wise interaction.   The strategy and procedure is applicable to   the more general cases when three-body forces are involved, due to the fact that the three-body dynamics and constraints on allowed energy spectra by periodic lattice structure are two independent procedures and can be carried out separately.

\section{ACKNOWLEDGMENTS}
  We   acknowledge support from Department of Physics and Engineering, California State University, Bakersfield, CA.

\appendix

\section{Three-body interaction  and Faddeev equations}\label{Faddeevpair}

In this section, we consider scattering of three spinless   particles of equal masses, interacting by $\delta$-function potentials of strength,     \mbox{$ V_{\alpha \beta}$}, between $\alpha$-th and $\beta$-th particles.   The  three-particle  wave function  satisfies Schr\"odinger  equation,
\begin{align}
& \left [- \frac{1}{2 m} \sum_{i=1}^{3} \frac{d^{2}}{d x_{i}^{2}}  + V_{12}\delta(r_{12}) + V_{23} \delta(r_{23}) + V_{31}\delta(r_{31})  - E \right ]  \nonumber \\
&\quad \quad  \times \Psi(x_{1}, x_{2}, x_{3}; p_{1},p_{2},p_{3})=0,
\end{align}
where   $m$,   $p_{i}$ (\mbox{$i=1,2,3$}) and  \mbox{$E= \sum_{i=1}^{3}\frac{p_{i}^{2}}{2m}$}  refer to    the mass of particle,    particle's inital momenta and three-body  total energy respectively. As shown in \cite{Guo:2016fgl},  the center of mass, relative  positions and corresponding conjugate momenta  among particles are defined by   \mbox{$R= \frac{ x_{1}+ x_{2}+x_{3}}{3}$}, \mbox{$r_{ij}=x_{i}-x_{j}$}    and \mbox{$r_{k}=\frac{x_{i} + x_{j}}{2} - x_{k}$},    \mbox{$P = p_{1} + p_{2}+p_{3}$}, \mbox{$q_{ij} = \frac{p_{i}-p_{j}}{2}$} and \mbox{$q_{k} = \frac{p_{i} + p_{j} - 2 p_{k}}{3}$} \mbox{($i \neq j \neq k$)} respectively. Because of translational invariance,   the center of mass motion is described by a plane wave,  the total three particles wave function is given by,  \mbox{$\Psi(x_{1}, x_{2},x_{3}; p_{1},p_{2},p_{3}) = e^{ i P R} \psi(r_{\alpha \beta}, r_{\gamma};q_{ij}, q_{k})$},  where \mbox{$\psi(r_{\alpha \beta}, r_{\gamma};q_{ij}, q_{k})$}   describes relative motions of three particles.

\subsection{Scattering of three free particles}  
 With an   incoming wave of   three free particles state,  \mbox{$\Psi_{(0)}$}, three-body  wave function has the  form \cite{Faddeev:1960su,Faddeev:1965},
  \mbox{$\Psi=\Psi_{(0)} +  \sum_{\gamma=1}^{3} \Psi_{(\gamma)}$},  
 where  $\Psi_{(\gamma)}$ satisfies equation,
\begin{align}
& \left [- \frac{1}{2 m} \sum_{i=1}^{3} \frac{d^{2}}{d x_{i}^{2}}  + V_{\alpha \beta} \delta(r_{\alpha \beta})   - E \right ] \Psi_{(\gamma)}  \nonumber \\
&\quad   \quad       = -V_{\alpha \beta}\delta(r_{\alpha \beta})  \left [ \Psi_{(0)}+ \Psi_{(\alpha)} + \Psi_{(\beta)}  \right ] , \ \ \ \gamma \neq \alpha \neq \beta. \label{3bpsik}
\end{align}
As shown in   \cite{Guo:2016fgl}, the Lippmann-Schwinger equation for relative wave function, $\psi_{(\gamma)} $,  can be obtained as
\begin{align}
& \psi_{(\gamma)}  (r_{\alpha \beta}, r_{\gamma};q_{ij}, q_{k})  =  \int_{-\infty}^{\infty}   \frac{d k}{2\pi}   e^{i \sqrt{\sigma^{2}    -\frac{3}{4} k^{2}} | r_{\alpha \beta} | }e^{i k  r_{\gamma}  }   \nonumber \\
&   \times      i t_{\alpha \beta}(\sqrt{\sigma^{2}   -\frac{3}{4} k^{2} } )   \int_{-\infty}^{\infty}  d r'_{\alpha \beta} d r'_{\gamma}    e^{ - i k r'_{\gamma} }    \delta(r'_{\alpha \beta}) \nonumber \\
&         \times \left [ \psi_{(0)}(r'_{\alpha \beta}, r'_{\gamma};q_{ij}, q_{k})   + \psi_{(\alpha)}  (r'_{\beta\gamma}, r'_{\alpha};q_{ij}, q_{k}) \right.  \nonumber \\
& \quad \quad  \quad \quad \quad      \quad  \quad    \quad       \quad    \quad     \left.  +  \psi_{(\beta)} (r'_{\gamma \alpha}, r'_{\beta};q_{ij}, q_{k})  \right ]    ,  \nonumber \\
& \quad \quad\quad \quad\quad \quad \quad \quad\quad \quad\quad  \alpha \neq \beta \neq \gamma,  \label{wavegamma}
\end{align} 
where     \mbox{$t_{\alpha \beta}(k) =- \frac{ m V_{\alpha \beta}}{2k+i m V_{\alpha \beta}} $}  refers to the two-body scattering amplitude between $\alpha$-th and $\beta$-th particles for a $\delta$-function potential interaction, and \mbox{$\sigma^{2}= m E - \frac{P^{2}}{6}=q_{ij}^{2} + \frac{3}{4} q^{2}_{k}$}.

Instead of solving Faddeev $T$-matrix equations, see \cite{Guo:2016fgl}, numerically, it is more convenient   to  introduce   amplitudes, $g_{(\gamma)}$, by
\begin{align}
& g_{(\gamma)}  (k ;  q_{ij}, q_{k})  = \int_{-\infty}^{\infty}   d r e^{-i k  r }   \nonumber \\
& \quad \times  \left [   \psi_{(\alpha)}  (r , - \frac{ r}{2};q_{ij}, q_{k})    +  \psi_{(\beta)} (r , - \frac{ r}{2};q_{ij}, q_{k})  \right ]  . \label{Ggammawav}
\end{align}
The   relative wave function, $\psi_{(\gamma)} $, thus can be written as
\begin{align}
& \psi_{(\gamma)}  (r_{\alpha \beta}, r_{\gamma};q_{ij}, q_{k})     \nonumber \\
&   =  \int_{-\infty}^{\infty}   \frac{d k}{2\pi}   e^{i \sqrt{\sigma^{2}    -\frac{3}{4} k^{2}} | r_{\alpha \beta} | }e^{i k  r_{\gamma}  }      i t_{\alpha \beta}(\sqrt{\sigma^{2}   -\frac{3}{4} k^{2} } )   \nonumber \\
&         \times \left [   \int_{-\infty}^{\infty}   d  r'_{\gamma}      e^{ - i k  r'_{\gamma}   } \psi_{(0)}(0,  r'_{\gamma}  ;q_{ij}, q_{k})   +g_{(\gamma)} (k ; q_{ij}, q_{k})   \right ]    .  \label{wavegammaG}
\end{align}
Using Eq.(\ref{Ggammawav}) and Eq.(\ref{wavegammaG}), the integral equations for  $g_{(\gamma)}$-amplitude are obtained,
\begin{align}
 g_{(\gamma)} (k ;   & q_{ij}, q_{k})    = g^{(0)}_{(\gamma)} (k ;   q_{ij}, q_{k})   
      \nonumber \\
  +  i  & \int_{-\infty}^{\infty}   \frac{d q}{2\pi}    \frac{2    \sqrt{\sigma^{2}    -\frac{3}{4} q^{2}}      }{ \left (\sigma^{2}    -\frac{3}{4} q^{2} \right ) - \left(k + \frac{q}{2} \right )^{2}  + i \epsilon }    \nonumber \\
&         \times  \left [   i t_{  \beta \gamma }(\sqrt{\sigma^{2}   -\frac{3}{4} q^{2} } )  g_{(\alpha)} (q ; q_{ij}, q_{k})  \right. \nonumber \\
 & \quad   \left. +  i t_{    \gamma \alpha }(\sqrt{\sigma^{2}   -\frac{3}{4} q^{2} } )  g_{(\beta)} (q ; q_{ij}, q_{k})   \right ]       , \label{faddeevGeq}
\end{align}
where $g^{(0)}_{(\gamma)}$   is given by,
\begin{align}
& g^{(0)}_{(\gamma)} (k ;   q_{ij}, q_{k})      =i  \int_{-\infty}^{\infty}   \frac{d q}{2\pi}   \frac{2    \sqrt{\sigma^{2}    -\frac{3}{4} q^{2}}      }{ \left (\sigma^{2}    -\frac{3}{4} q^{2} \right ) - \left(k + \frac{q}{2} \right )^{2}  + i \epsilon }      \nonumber \\
&         \times \left [   i t_{ \beta \gamma}(\sqrt{\sigma^{2}   -\frac{3}{4} q^{2} } )  \int_{-\infty}^{\infty}   d r_{\alpha}    e^{ - i q r_{\alpha} } \psi_{(0)}(0, r_{\alpha};q_{ij}, q_{k}) \right.   \nonumber \\
& \quad \left. + i t_{\gamma \alpha}(\sqrt{\sigma^{2}   -\frac{3}{4} q^{2} } )  \int_{-\infty}^{\infty}   d r_{\beta}    e^{ - i q r_{\beta} } \psi_{(0)}(0, r_{\beta};q_{ij}, q_{k})   \right ]    .   \label{g0}
\end{align}

The $g_{(\gamma)}$-amplitude is related to standard Faddeev $T$-amplitude by
\begin{align}
& T_{(\gamma)}  (k ;  q_{ij}, q_{k}) =  2\sqrt{\sigma^{2}   -\frac{3}{4} k^{2} }     t_{\alpha \beta}(\sqrt{\sigma^{2}   -\frac{3}{4} k^{2} } )   \nonumber \\
& \quad \times \left [    \int_{-\infty}^{\infty}    d r_{\gamma}  e^{-i k  r_{\gamma}}     \psi_{(0)}  (0, r_{\gamma}; q_{ij}, q_{k})      + g_{(\gamma)} (k ;   q_{ij}, q_{k})    \right ], \label{gTrelation}
\end{align}
the full three-body scattering amplitude is given by 
 \begin{align}
T  (k_{\alpha \beta}, k_{\gamma}; q_{ij}, q_{k}) =  \sum_{\delta =1}^{3}T_{(\delta)} (k_{\delta}; q_{ij}, q_{k})  ,  \label{3bTsum}
\end{align}
where \mbox{$k_{\alpha} =- k_{\alpha \beta} -\frac{k_{\gamma}}{2}$} and \mbox{$k_{\beta} = k_{\alpha \beta} -\frac{k_{\gamma}}{2}$}.

\subsection{Scattering on a two-body bound state}  
For the case of scattering on a bound state, {\it e.g.} i-th particle incident on a bound state of  $(jk)$ pair, three-body  wave function  thus has the form of   \mbox{$\Psi=   \sum_{\gamma=1}^{3} \Psi_{(\gamma)}$},   and  the Lippmann-Schwinger equation for relative wave function reads,
\begin{align}
 \psi_{(\gamma)}  & (r_{\alpha \beta}, r_{\gamma};q_{ij}, q_{k})  = \delta_{\gamma, i} \phi^{B}_{(\gamma)}(r_{\alpha \beta}) e^{i q_{i}^{B} r_{\gamma}}  \nonumber \\
& +  \int_{-\infty}^{\infty}   \frac{d k}{2\pi}   e^{i \sqrt{\sigma^{2}    -\frac{3}{4} k^{2}} | r_{\alpha \beta} | }e^{i k  r_{\gamma}  }   \nonumber \\
&  \quad \quad  \times      i t_{\alpha \beta}(\sqrt{\sigma^{2}   -\frac{3}{4} k^{2} } )   g_{(\gamma)} (k ; q_{ij}, q_{k})   , \label{wavegammabound}
\end{align}
where     the bound state wave function in pair $( j k)$ is given by \mbox{$\phi^{B}_{(\gamma)}(r_{\alpha \beta})  = \sqrt{ - \frac{m V_{j k}}{2}} e^{ \frac{m V_{j k}}{2} |r_{\alpha \beta}| }$},  and \mbox{$q_{i}^{B} = \sqrt{ \sigma^{2} + \frac{3}{4} (\frac{m V_{jk}}{2})^{2} }$} refers to the relative momentum between incident i-th particle and pair $(jk)$ bound state.
The integral equations for  $g_{(\gamma)}$-amplitude for scattering on a bound state     also has the same form of Eq.(\ref{faddeevGeq}),
where for the case of scattering on a bound state in  $(jk)$ pair,  $g_{(\gamma)}^{(0)}$   is given by
\begin{align}
 g^{(0)}_{(\gamma)} (k ;   q_{ij}, q_{k})    &   =  \delta_{\alpha, i}  \int d r_{\alpha}  \phi^{B}_{(\alpha)}( r_{\alpha}) e^{ - i  ( k+\frac{ q_{i}^{B} }{2} ) r_{\alpha}}   \nonumber \\
 & +      \delta_{\beta, i}  \int d r_{\beta}  \phi^{B}_{(\beta)}( r_{\beta}) e^{ - i  ( k+\frac{ q_{i}^{B} }{2} ) r_{\beta}}    .  
\end{align}

\subsection{Exact solutions for equal strength of $\delta$-function potentials: $V_{12}=V_{23}=V_{31}=V_{0}$}

Only for the special case with equal strength of $\delta$-potential among all pairs, $V_{12}= V_{23}= V_{31} = V_{0}$, three-body interactions    are in fact exactly solvable  \cite{Guo:2016fgl,Dodd:1970,Majumdar:1972}. The exact solutions can be used for testing and verifying numerical approach, and also for completeness, the exact solutions of Faddeev equations, Eq.(\ref{faddeevGeq}),  are presented in follows for scattering of three free particles with incoming wave $e^{i q_{12} r_{12}} e^{i q_{3} r_{3}}$ and for scattering on a bound state of pair $(12)$ respectively.

\subsubsection{Scattering of three free particles with incoming wave $e^{i q_{12} r_{12}} e^{i q_{3} r_{3}}$}\label{exactgsol}
For incoming wave $e^{i q_{12} r_{12}} e^{i q_{3} r_{3}}$,   $g_{(\gamma)}^{(0)}$ are thus given by
\begin{align}
 g_{(3)}^{(0)} (k;q_{ij}, q_{k}) & = i \frac{2 q_{23} i t (- q_{23})}{(k- q_{3} - i \epsilon) (k - q_{2} - i \epsilon)} \nonumber \\
 &+  i \frac{2 q_{31} i t  (- q_{31})}{(k- q_{3} - i \epsilon) (k - q_{1} + i \epsilon)}  \nonumber \\
 & + it  (-q_{23}) 2 \pi \delta(k- q_{3})  ,  \label{g0free1}
 \end{align}
 \begin{align}
   g_{(1)}^{(0)} (k;q_{ij}, q_{k})   &=  i \frac{2 q_{31} i t  (- q_{31})}{(k- q_{3} - i \epsilon) (k - q_{1} + i \epsilon)}   \nonumber \\
 & - i \frac{2 q_{12} i t  ( q_{12})}{(k- q_{2} - i \epsilon) (k - q_{1} + i \epsilon)}  , \label{g0free2}
  \end{align}
 \begin{align}
  g_{(2)}^{(0)} (k;q_{ij}, q_{k})  &= - i \frac{2 q_{12} i t  ( q_{12})}{(k- q_{2} - i \epsilon) (k - q_{1} + i \epsilon)} \nonumber \\
 &+ i \frac{2 q_{23} i t  (- q_{23})}{(k- q_{3} - i \epsilon) (k - q_{2} - i \epsilon)} \nonumber \\
 & + it  (-q_{23}) 2 \pi \delta(k- q_{3})  , \label{g0free3}
\end{align}
where \mbox{$t(q) =- \frac{m V_{0}}{2q + i m V_{0}}$}.  The exact solutions of $g$-amplitude are
\begin{align}
 g_{(3)} (k  & ;q_{ij}, q_{k})  =  it (-q_{23}) 2 \pi \delta(k- q_{3})  \nonumber \\
&+  \frac{\left (1+ \frac{  \frac{i m V_{0}}{2}}{ \sqrt{\sigma^{2} - \frac{3}{4} k^{2}} } \right )  \frac{  (-2  m V_{0}) (k + \frac{q_{3}}{2}) }{ \left (1+ \frac{ i m V_{0}}{2 q_{12}} \right )   \left (1-  \frac{ i m V_{0}}{2 q_{23}} \right )  \left (1- \frac{ i m V_{0}}{2 q_{31}} \right )  } }{  (k- q_{3} - i \epsilon)(k- q_{2} - i \epsilon) (k - q_{1} + i \epsilon)}, \label{g3exact}
\end{align}
\begin{align}
 g_{(1)} (k & ;q_{ij}, q_{k})    \nonumber \\
 &=  \frac{\left (1+ \frac{  \frac{i m V_{0}}{2}}{ \sqrt{\sigma^{2} - \frac{3}{4} k^{2}} } \right )  \frac{  (-2  m V_{0}) (k -\frac{q_{12}}{2}  - \frac{q_{3}}{4} -\frac{i m V_{0}}{2} ) }{ \left (1+ \frac{ i m V_{0}}{2 q_{12}} \right )   \left (1-  \frac{ i m V_{0}}{2 q_{23}} \right )  \left (1- \frac{ i m V_{0}}{2 q_{31}} \right )  } }{  (k  -  q_{3} - i \epsilon)(k- q_{2} - i \epsilon) (k - q_{1} + i \epsilon)}, \label{g1exact}
\end{align}
 \begin{align}
 g_{(2)} (k &; q_{ij}, q_{k}) =  it  (-q_{23}) 2 \pi \delta(k- q_{3})  \nonumber \\
&+  \frac{\left (1+ \frac{  \frac{i m V_{0}}{2}}{ \sqrt{\sigma^{2} - \frac{3}{4} k^{2}} } \right )  \frac{  (-2  m V_{0}) (k +\frac{q_{12}}{2}  - \frac{q_{3}}{4} +\frac{i m V_{0}}{2}  ) }{ \left (1+ \frac{ i m V_{0}}{2 q_{12}} \right )   \left (1-  \frac{ i m V_{0}}{2 q_{23}} \right )  \left (1- \frac{ i m V_{0}}{2 q_{31}} \right )  } }{  (k- q_{3} - i \epsilon)(k- q_{2} - i \epsilon) (k - q_{1} + i \epsilon)}.\label{g2exact}
\end{align}

\subsubsection{Scattering on a bound state of pair $(12)$}
For incoming wave $ \phi^{B}_{(3)}(r_{12}) e^{i q_{3}^{B} r_{3}} $, the $g_{(\gamma)}^{(0)}$ are given by
\begin{align}
& g_{(3)}^{(0)} (k;q_{ij}, q_{k})  =  0 ,   \ \ \  \nonumber \\
&   g_{(1,2)}^{(0)} (k;q_{ij}, q_{k})     =   \frac{2\sqrt{ (\frac{ m V_{0}}{2})^{2} }  \sqrt{ - \frac{m V_{0}}{2}}  }{  \left (k + \frac{q_{3}^{B}}{2} \right )^{2} + ( \frac{m V_{0}}{2} )^{2} } .   \label{g0bound}
  \end{align}
 The exact solutions of $g$-amplitude are
\begin{align}
 g_{(3)} (k & ; q_{ij}, q_{k} )  = \frac{4}{3}\sqrt{ (\frac{ m V_{0}}{2})^{2} }  \sqrt{ - \frac{m V_{0}}{2}}   \nonumber \\
 & \times   \frac{  \left (1+ \frac{  \frac{i m V_{0}}{2}}{ \sqrt{\sigma^{2} - \frac{3}{4} k^{2}} } \right )   \frac{  \left ( \frac{3 q_{3}}{2}  -  \frac{ i m V_{0}}{2  } \right )    }{   \left ( \frac{q_{3}}{2}  +  \frac{ i m V_{0}}{2  } \right )  }  (k + \frac{q_{3}}{2}  + \frac{i m V_{0}}{2} ) }{  (k- q_{3} - i \epsilon)(k- q_{2} - i \epsilon) (k - q_{1} + i \epsilon)},
\end{align}
\begin{align}
 g_{(1,2)} (k & ; q_{ij}, q_{k} )   = \frac{4}{3}\sqrt{ (\frac{ m V_{0}}{2})^{2} }  \sqrt{ - \frac{m V_{0}}{2}}   \nonumber \\
 & \times   \frac{  \left (1+ \frac{  \frac{i m V_{0}}{2}}{ \sqrt{\sigma^{2} - \frac{3}{4} k^{2}} } \right )   \frac{  \left ( \frac{3 q_{3}}{2}  -  \frac{ i m V_{0}}{2  } \right )    }{   \left ( \frac{q_{3}}{2}  +  \frac{ i m V_{0}}{2  } \right )  }  (k - \frac{q_{3}}{4}  - \frac{i m V_{0}}{4} ) }{  (k- q_{3} - i \epsilon)(k- q_{2} - i \epsilon) (k - q_{1} + i \epsilon)}.
\end{align}

\subsection{Numerical test for scattering of three free particles with an  incoming wave $e^{i q_{12} r_{12}} e^{i q_{3} r_{3}}$}

  \begin{figure}
\begin{center}
\includegraphics[width=3.2 in,angle=0]{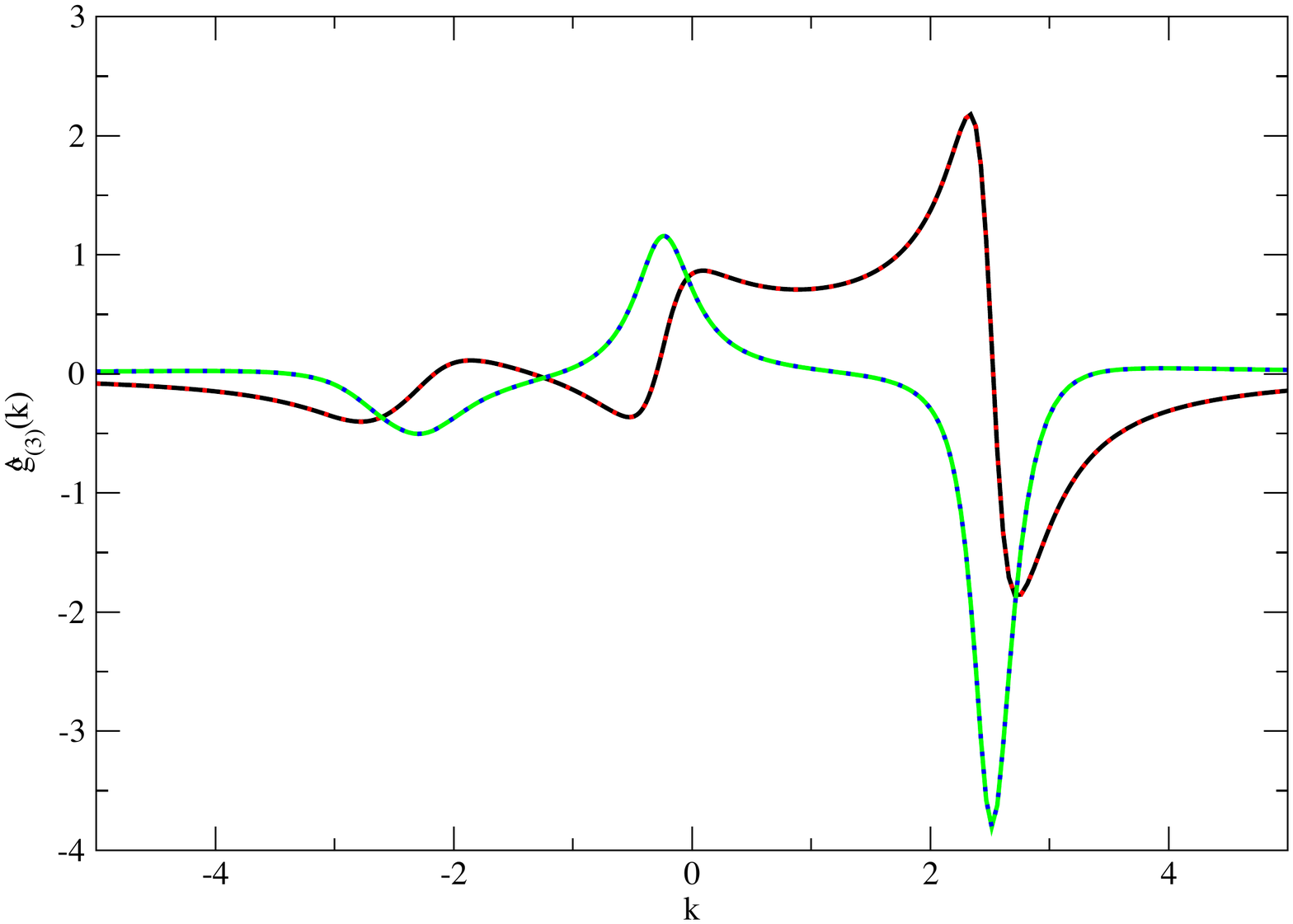}  
\includegraphics[width=3.2 in,angle=0]{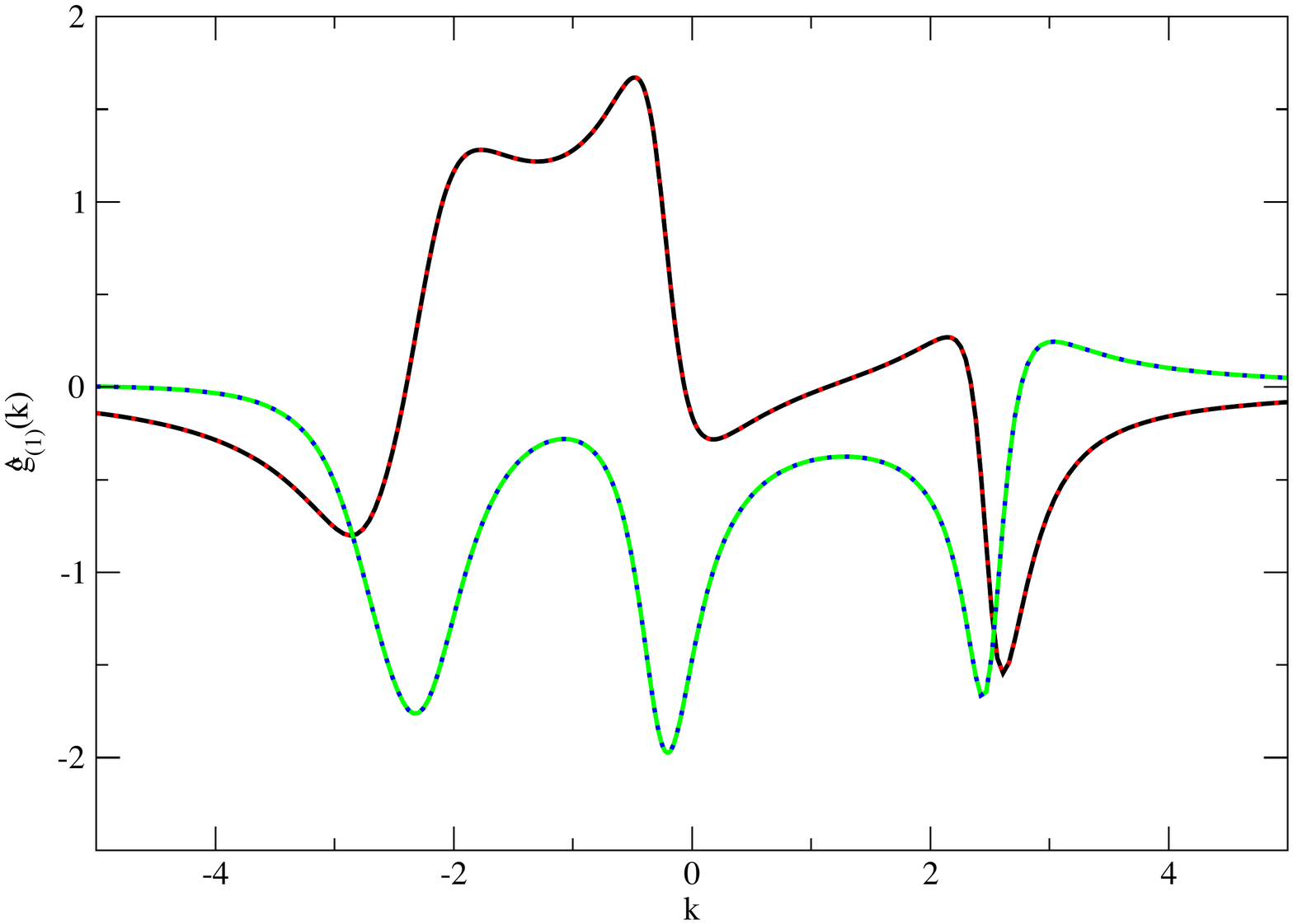} 
\includegraphics[width=3.2 in,angle=0]{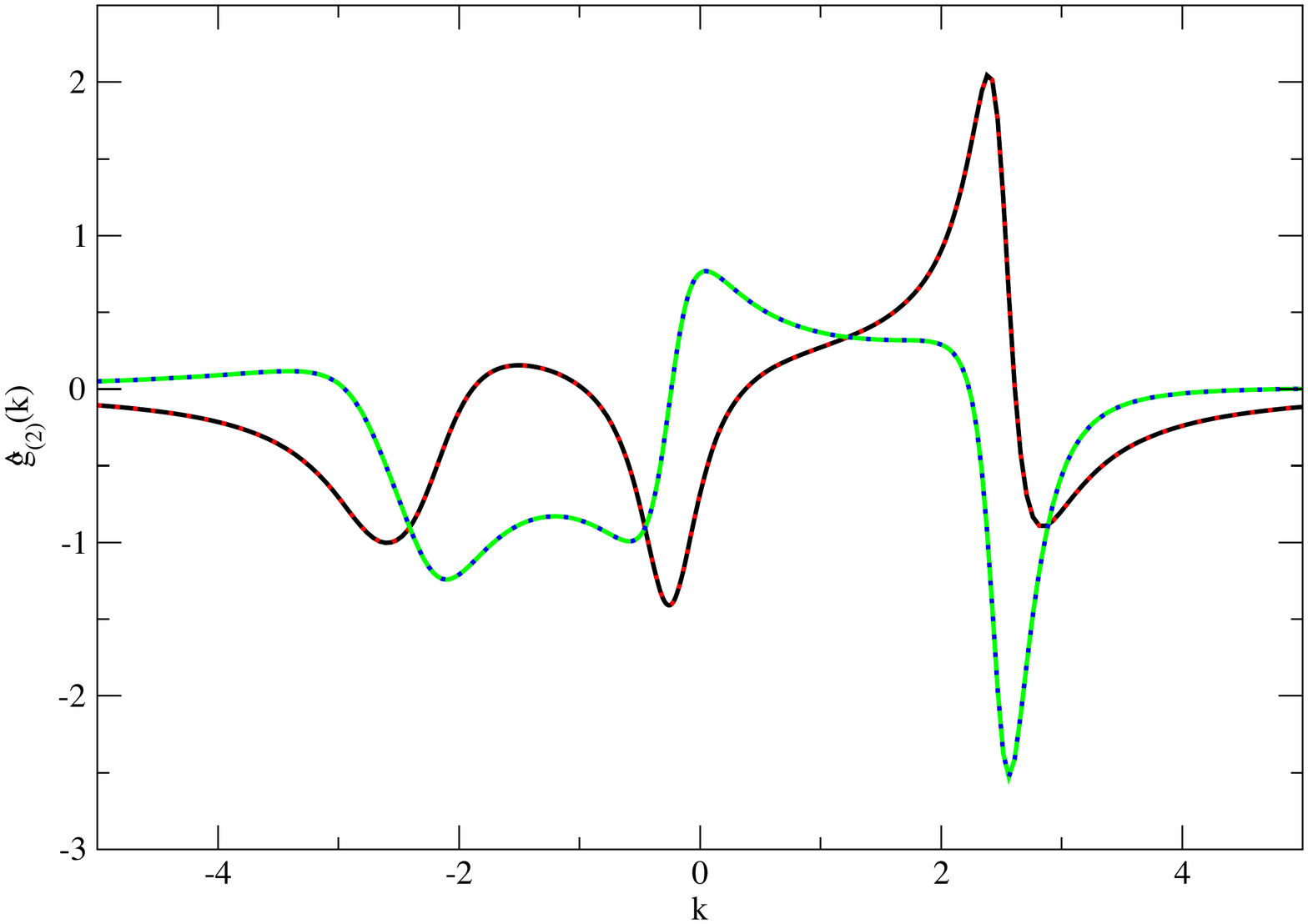} 
\caption{ The comparison of numerical solutions of  integral Eq.(\ref{gfreefredholm}) with exact solutions given in Eqs.(\ref{g3exact}-\ref{g2exact}):  the solutions of $\hat{g}_{(3,1,2)}$ are presented in upper, middle and lower panels respectively.  Solid black and solid green curves represent real and imaginary parts of numerical solutions, and dotted red and blue are real and imaginary parts of exact solutions respectively. The parameters of the toy model are choosen as $mV_{0}=2.0$, $q_{12} =1.0+0.4 i$ and $q_{3} =2.5+ 0.2 i$, where  an imaginary part is given to both $(q_{12},q_{3})$ to smooth out the curves near the pole position for a better visualization purpose only.   \label{fig:gs}  } 
\end{center}
\end{figure}

In general, Faddeev equations, Eq.(\ref{faddeevGeq}), have to be solved numerically. The numerical approach is rather straightforward for the case of scattering on a bound state,    the expression of $g_{(\gamma)}^{(0)}$ (see Eq.(\ref{g0bound}))  does not contain  $\delta$-function type  singularities.  Thus Eq.(\ref{faddeevGeq}) is standard Fredholm-type integral equation, and can be solved easily by matrix inversion method. The special care has to be given to the case of scattering of three free particles, in this case,  $g_{(\gamma)}^{(0)}$ (see Eq.(\ref{g0free1}-\ref{g0free3})) does indeed contain $\delta$-function type singularities. The singularities in Eq.(\ref{faddeevGeq})   can be removed by redefining $g_{(\gamma)}$'s. For example,  given $g_{(3)}^{(0)} $ contains singular term $it_{23} (-q_{23}) 2 \pi \delta(k- q_{3}) $,  by a shifting in $g_{(3)}$,
\begin{align}
g_{(3)} (k;q_{ij}, q_{k}) = \hat{g}_{(3)}  (k;q_{ij}, q_{k})+ it_{23} (-q_{23}) 2 \pi \delta(k- q_{3}) ,
\end{align}
the new integral equations for $ \hat{g}_{(\gamma)} $'s are thus free of $\delta$-function type singularities, and  are also Fredholm-type   equations. In addition, extra care   has to be taken when it comes to the branch cut of square root terms, and pole contributions in Faddeev equations. The  pole contributions are handled by using standard $i \epsilon$ prescription, see Eq.(\ref{faddeevGeq}) for instance. As for branch cut contribution, we adopt the same convention as used in \cite{Guo:2016fgl}, for the square root terms,  $\sqrt{q_{\alpha \beta}^{2}}$,  we  assign a small imaginary part to \mbox{$q_{12} \rightarrow q_{12} + i 0^{+}$},   the imaginary part for \mbox{$q_{23} \rightarrow q_{23} - i 0^{+}$} and \mbox{$q_{31}\rightarrow q_{31} - i 0^{+}$} are thus determined   by relations, \mbox{$q_{23} = -\frac{1}{2}q_{12} + \frac{3}{4} q_{3}$} and \mbox{$q_{31} = -\frac{1}{2}q_{12} - \frac{3}{4} q_{3}$} respectively.   In addition, our convention for complex square root is given by $\sqrt{q^{2} \pm i 0^{+}} = \pm \sqrt{q^{2}}$, therefore, \mbox{$\sqrt{ (q_{12}+ i 0^{+})^{2}} = q_{12}$}, \mbox{$\sqrt{ (q_{23}- i 0^{+})^{2}} = -q_{23}$} and \mbox{$\sqrt{ (q_{31} - i 0^{+})^{2}} = - q_{31}$}.

As the demonstrations of some numerical tests, the $g_{(\gamma)}$'s equations are solved numerically for an incoming wave of $e^{i q_{12} r_{12}} e^{i q_{3} r_{3}}$ and compared with the exact solutions  presented in section \ref{exactgsol}. As already mentioned  previously, the $\delta$-function type singularities have to be removed by shifting $g_{(\gamma)}$'s,
\begin{align}
g_{(1)} (k;q_{ij}, q_{k})  &= \hat{g}_{(1)}  (k;q_{ij}, q_{k}), \nonumber \\
g_{(3,2)} (k;q_{ij}, q_{k}) &= \hat{g}_{(3,2)}  (k;q_{ij}, q_{k})+ it_{23} (-q_{23}) 2 \pi \delta(k- q_{3}), 
\end{align} 
we thus obtain integral equations for $\hat{g}_{(\gamma)}$'s,
 \begin{align}
&  \hat{G}(k  )   
=  \hat{G}^{(0)}(k )     \nonumber \\
&+ i \int_{-\infty}^{\infty}   \frac{d q}{2\pi}    \frac{2    \sqrt{\sigma^{2}    -\frac{3}{4} q^{2}}      }{ \sigma^{2}    -\frac{3}{4} q^{2}  - \left( k + \frac{q}{2} \right)^{2}  + i \epsilon }     \mathcal{K}(\sqrt{\sigma^{2}   -\frac{3}{4} q^{2} } )   \hat{G}(q  )   ,\label{gfreefredholm}
 \end{align}
 where $\hat{G}$ and  $\hat{G}^{(0)}$ stand for column vectors $(  \hat{ g}_{(3)} ,  \hat{ g}_{(1)}, \hat{ g}_{(2)} )^{T}$ and $(  \hat{ g}^{(0)}_{(3)} ,  \hat{ g}^{(0)}_{(1)}, \hat{ g}^{(0)}_{(2)} )^{T}$ respectively, the dependence on initial momenta $(q_{ij}, q_{k})$ has been dropped in above equation.    The   matrix $ \mathcal{K}$ is given by
  \begin{align}
&  \mathcal{K}( q) 
=    \begin{bmatrix}
0 &  i t_{  23 } (q) & i t_{ 31} (q)  \\
 it_{12}(q) & 0 & it_{31} (q)    \\
it_{12} (q) & it_{23} (q) & 0
\end{bmatrix}    , \label{gkern}
 \end{align}
and
\begin{align}
\hat{ g}_{(3)}^{(0)} (k;q_{ij}, q_{k}) & = i \frac{2 q_{23} i t_{23} (- q_{23})}{(k- q_{3} - i \epsilon) (k - q_{2} - i \epsilon)} \nonumber \\
 & +  i \frac{2 q_{31} i t_{31} (- q_{31})}{(k- q_{3} - i \epsilon) (k - q_{1} + i \epsilon)}  \nonumber \\
 & - i \frac{2 q_{12} i t_{31} ( q_{12}) i t_{23} (- q_{23})}{(k- q_{2} - i \epsilon) (k - q_{1} + i \epsilon)}  ,   
 \end{align}
 \begin{align}
   \hat{g}_{(1)}^{(0)} (k;q_{ij}, q_{k})   &=  i \frac{2 q_{31} i t_{31} (- q_{31})}{(k- q_{3} - i \epsilon) (k - q_{1} + i \epsilon)}   \nonumber \\
 & - i \frac{2 q_{12} i t_{12} ( q_{12})}{(k- q_{2} - i \epsilon) (k - q_{1} + i \epsilon)}  ,  
  \end{align}
 \begin{align}
\hat{  g}_{(2)}^{(0)} (k;q_{ij}, q_{k})  &= - i \frac{2 q_{12} i t_{12} ( q_{12}) \left [ 1+ i t_{23} (- q_{23}) \right ]}{(k- q_{2} - i \epsilon) (k - q_{1} + i \epsilon)} \nonumber \\
 &+ i \frac{2 q_{23} i t_{23} (- q_{23})}{(k- q_{3} - i \epsilon) (k - q_{2} - i \epsilon)}   . 
\end{align}

Faddeev equations for $\hat{g}_{(\gamma)} $'s, given  by Eq.(\ref{gfreefredholm}),  are   solved numerically by matrix inversion method, and the comparison of numerical solutions  with exact solutions in  Eqs.(\ref{g3exact}-\ref{g2exact}) is presented in Fig.\ref{fig:gs}.

\section{Faddeev equations including three-body force}\label{Faddeev3bforce}
In previous sections,  our discussion  of three-body problem has been restricted on the  interactios of three particles with only pair-wise $\delta$-function potentials. In this section, we would like to extend our discussion of three-body interaction by including a spherical symmetric  three-body force potential, $U(r )$ where \mbox{$r=\sqrt{r_{\alpha\beta}^{2} + \frac{4}{3} r_{\gamma}^{2}}$}, and give a brief presentation how the three-body force may be handled in Faddeev equations approach. By including a three-body force potential, $U(r )$,   Schr\"odinger  equation now has the form of
\begin{align}
& \left [- \frac{1}{2 m} \sum_{i=1}^{3} \frac{d^{2}}{d x_{i}^{2}}  + \sum_{\gamma=1}^{3} V_{\alpha \beta}\delta(r_{\alpha \beta})  + U(r )  - E \right ]  \nonumber \\
&\quad \quad  \times \Psi(x_{1}, x_{2}, x_{3}; p_{1},p_{2},p_{3})=0.
\end{align}
Let's consider the scattering of three-particle with an incoming wave of  three free particles, $\Psi_{(0)}$, three-body wave function may thus be expressed in the form of 
\begin{align}
\Psi = \Psi_{(0)} + \sum_{\gamma=1}^{3} \Psi_{(\gamma)} + \Psi_{(U)},
\end{align}
 where  $\Psi_{(\gamma)}$  satisfies equation,
\begin{align}
& \left [- \frac{1}{2 m} \sum_{i=1}^{3} \frac{d^{2}}{d x_{i}^{2}}  + V_{\alpha \beta} \delta(r_{\alpha \beta})   - E \right ] \Psi_{(\gamma)}  \nonumber \\
&\quad   \quad       = -V_{\alpha \beta}\delta(r_{\alpha \beta})  \left [ \Psi_{(0)}+ \Psi_{(\alpha)} + \Psi_{(\beta)} + \Psi_{(U)}  \right ] , \nonumber \\
& \ \ \ \   \ \ \ \ \ \ \   \ \ \ \ \ \ \   \ \ \ \ \ \ \   \ \ \ \gamma \neq \alpha \neq \beta,  
\end{align}
and similarly the equation for $\Psi_{(U)}$ is given by
\begin{align}
& \left [- \frac{1}{2 m} \sum_{i=1}^{3} \frac{d^{2}}{d x_{i}^{2}}  + U(r )  - E \right ] \Psi_{(U)}  \nonumber \\
&\quad   \quad       = - U(r )  \left [ \Psi_{(0)}+ \Psi_{(\alpha)} + \Psi_{(\beta)} + \Psi_{(\gamma)}  \right ] .
\end{align}

The Lippmann-Schwinger equation for relative wave function, $\psi_{(\gamma)} $ and $\psi_{(U)}$,  can be obtained respectively as
\begin{align}
& \psi_{(\gamma)}  (r_{\alpha \beta}, r_{\gamma};q_{ij}, q_{k})  =  \int_{-\infty}^{\infty}   \frac{d k}{2\pi}   e^{i \sqrt{\sigma^{2}    -\frac{3}{4} k^{2}} | r_{\alpha \beta} | }e^{i k  r_{\gamma}  }   \nonumber \\
&   \times      i t_{\alpha \beta}(\sqrt{\sigma^{2}   -\frac{3}{4} k^{2} } )   \int_{-\infty}^{\infty}  d r'_{\alpha \beta} d r'_{\gamma}    e^{ - i k r'_{\gamma} }    \delta(r'_{\alpha \beta}) \nonumber \\
&         \times \left [ \psi_{(0)}(r'_{\alpha \beta}, r'_{\gamma};q_{ij}, q_{k})   + \psi_{(\alpha)}  (r'_{\beta\gamma}, r'_{\alpha};q_{ij}, q_{k}) \right.  \nonumber \\
& \quad       \left.  +  \psi_{(\beta)} (r'_{\gamma \alpha}, r'_{\beta};q_{ij}, q_{k}) +  \psi_{(U)} (r'_{ \alpha \beta}, r'_{\gamma};q_{ij}, q_{k})  \right ]     ,  \nonumber \\
& \quad \quad\quad \quad\quad \quad \quad  \quad \quad \quad \quad \quad \quad   \alpha \neq \beta \neq \gamma,
\end{align} 
 and
 \begin{align}
& \psi_{(U)}  (r_{12}, r_{3};q_{ij}, q_{k})   \nonumber \\
& =  \int_{- \infty}^{\infty} d r'_{12}  d r'_{3} G_{(U)} (r_{12}, r_{3}; r'_{12}, r'_{3}; \sigma) m U(r' ) \nonumber \\
&         \times \left [ \psi_{(0)}(r'_{12}, r'_{3};q_{ij}, q_{k})   +  \sum_{\gamma=1}^{3}  \psi_{(\gamma)} (r'_{ \alpha \beta}, r'_{\gamma};q_{ij}, q_{k})  \right ]   ,  \nonumber \\
& \quad \quad\quad \quad\quad \quad \quad  \quad \quad \quad \quad \quad \quad    \alpha \neq \beta \neq \gamma  .
\end{align} 
The Green's function, $G_{(U)}$,  satisfies  equation,
\begin{align}
& \left [  \sigma^{2}  + \frac{d^{2}}{d r^{2}_{12}} + \frac{3}{4} \frac{d^{2}}{d r^{2}_{3}}  - m U(r )    \right ]    G_{(U)} (r_{12}, r_{3}; r'_{12}, r'_{3}; \sigma)  \nonumber \\
& \quad \quad \quad \quad  \quad \quad \quad \quad  =   \delta(r_{12} - r'_{12})  \delta(r_{3} -r'_{3}) .
\end{align}

Next, let's introduce the scattering  amplitudes by
\begin{align}
 &g_{(\gamma)}(k; q_{ij}, q_{k}) =  \int_{-\infty}^{\infty}   d r e^{-i k  r }    \nonumber \\
& \quad  \times  \left[   \psi_{(\alpha)}  (r , - \frac{ r}{2};q_{ij}, q_{k})    +  \psi_{(\beta)} (r , - \frac{ r}{2};q_{ij}, q_{k})   \right.   \nonumber \\
& \quad \quad\quad \quad  \quad \quad\quad \quad  \quad \quad  \quad  \left.  + \psi_{(U)}  (0, r  ;q_{ij}, q_{k})   \right], \label{ggamma3bU} \\
&T_{(\gamma)}(k; q_{ij}, q_{k}) = -  \int_{- \infty}^{\infty} d r_{\alpha \beta} d r_{\gamma}   e^{-i k   r_{\gamma}}  \nonumber \\
 &  \quad \quad\quad \quad \quad \quad\times m V_{\alpha \beta} \delta(r_{\alpha\beta})  \psi (r_{\alpha \beta}, r_{\gamma}; q_{ij},q_{k}), \label{Tgamma3bU} \\
 &T_{(U)} (k_{12}, k_{3}; q_{ij}, q_{k})= -  \int_{- \infty}^{\infty} d r_{12} d r_{3}    e^{-i k_{12} r_{12}}  e^{-i k_{3}  r_{3} }     \nonumber \\
 &  \quad \quad\quad \quad\quad \quad  \times m U(r )  \psi (r_{12}, r_{3}; q_{ij},q_{k}). \label{TU3bU}
\end{align}
The    $T_{(\gamma)}$- and $g_{(\gamma)}$-amplitudes are still related by Eq.(\ref{gTrelation}), and the total three-body scattering amplitude with the presence of three-body force   is thus given by
\begin{align}
& T (k_{12}, k_{3}; q_{ij}, q_{k})   \nonumber \\
& \quad = \sum_{\gamma=1}^{3} T_{(\gamma)}(k_{\gamma}; q_{ij}, q_{k}) + T_{(U)} (k_{12}, k_{3}; q_{ij}, q_{k}), 
\end{align}
where $k_{\alpha } = - k_{\alpha \beta} - \frac{k_{\gamma}}{2}$ and $k_{\beta } =  k_{\alpha \beta} - \frac{k_{\gamma}}{2}$. 

The wave functions, $\psi_{(\gamma)}$ and $\psi_{(U)}$, are given in terms of $T$-amplitude by
    \begin{align}
 &  \psi_{(\gamma)}(r_{\alpha \beta}, r_{\gamma}; q_{ij},q_{k})   \nonumber \\
 &     =  i \int_{- \infty}^{\infty} \frac{d k}{2\pi}  \frac{e^{i \sqrt{\sigma^{2} - \frac{3}{4} k^{2}} |r_{\alpha \beta}| } e^{i k r_{\gamma }} }{  2 \sqrt{\sigma^{2} - \frac{3}{4} k^{2}}     } T_{(\gamma)}(k; q_{ij}, q_{k})  , \label{wavgamma3bU}  \\
 & \psi_{(U)}(r_{12}, r_{3}; q_{ij},q_{k})   \nonumber \\
 &= - \int_{- \infty}^{\infty} \frac{d k_{12}}{2\pi}  \frac{d k_{3}}{2\pi} \frac{e^{i k_{12} r_{12}}  e^{i k_{3}  r_{3} } T_{(U)} (k_{12}, k_{3}; q_{ij}, q_{k})  }{\sigma^{2} - k_{12}^{2} - \frac{3}{4} k_{3}^{2} + i \epsilon}  . \label{waveU3bU}
\end{align}

Eq.(\ref{Tgamma3bU}-\ref{waveU3bU})  yield a   sets of coupled equations for $g_{(\gamma)}$ and $T_{(U)}$ amplitudes respectively, 
\begin{align}
 g_{(\gamma)} (k ;   & q_{ij}, q_{k})    = g^{(0)}_{(\gamma)} (k ;   q_{ij}, q_{k})    \nonumber \\
      - &  \int_{- \infty}^{\infty} \frac{d q}{2\pi}   \frac{  T_{(U)} (q, k; q_{ij}, q_{k})  }{\sigma^{2} - q^{2} - \frac{3}{4} k^{2} + i \epsilon}  \nonumber \\
  +   & i \int_{-\infty}^{\infty}   \frac{d q}{2\pi}    \frac{2    \sqrt{\sigma^{2}    -\frac{3}{4} q^{2}}      }{ \left (\sigma^{2}    -\frac{3}{4} q^{2} \right ) - \left(k + \frac{q}{2} \right )^{2}  + i \epsilon }    \nonumber \\
&         \times  \left [   i t_{  \beta \gamma }(\sqrt{\sigma^{2}   -\frac{3}{4} q^{2} } )  g_{(\alpha)} (q ; q_{ij}, q_{k})  \right. \nonumber \\
 & \quad   \left. +  i t_{    \gamma \alpha }(\sqrt{\sigma^{2}   -\frac{3}{4} q^{2} } )  g_{(\beta)} (q ; q_{ij}, q_{k})   \right ]       ,  \nonumber \\
 &  \quad \quad  \quad\quad  \quad \quad  \quad\quad   \quad\quad     \alpha \neq \beta \neq \gamma ,
\end{align}
where $g^{(0)}_{(\gamma)}$   is defined in Eq.(\ref{g0}), and
 \begin{align}
& T_{(U)} (k_{12},k_{3}; q_{ij}, q_{k}) = v_{(U)} (k_{12},k_{3}; q_{ij}, q_{k}) \nonumber \\
& \quad \quad  +   \sum_{\gamma=1}^{3}  \int_{-\infty}^{\infty} \frac{d q}{2\pi} \mathcal{F}_{(\alpha \beta)} (k_{12},k_{3};q )   \nonumber \\
& \quad \quad\quad    \times     i t_{\alpha \beta}(\sqrt{\sigma^{2}   -\frac{3}{4} q^{2} } )  g_{(\gamma)}  ( q ;q_{ij}, q_{k})   , \nonumber \\
&  \quad \quad  \quad\quad    \quad  \quad\quad   \quad\quad \quad \quad   \quad\quad   \alpha \neq \beta \neq \gamma .
\end{align}
The functions,  $v_{(U)}$ and $ \mathcal{F}_{(\alpha \beta)}$, are defined respectively by
\begin{align}
& v_{(U)} (k_{12},k_{3}; q_{ij}, q_{k})  = -  \int_{-\infty}^{\infty}  d r_{12} d r_{3}     \nonumber \\
&  \quad   \times  \left [  \psi_{(0)} (r_{12}, r_{3};q_{ij}, q_{k}) + \sum_{\gamma=1}^{3}  \psi^{(in)}_{(\gamma)} (r_{\alpha \beta}, r_{\gamma};q_{ij}, q_{k})  \right ]  \nonumber \\
& \quad \times m U(r )   \phi^{*}_{(U)}  (r_{12}, r_{3}; k_{12}, k_{3}) ,
\end{align}
and
\begin{align}
 \mathcal{F}_{(\alpha \beta)} (k_{12},k_{3};q )  
&= -  \int_{-\infty}^{\infty}  d r_{12} d r_{3}   e^{i \sqrt{\sigma^{2} - \frac{3}{4} q^{2}} |r_{\alpha \beta}| } e^{i q r_{\gamma }}  \nonumber \\
&   \times  m U(r ) \phi^{*}_{(U)}  (r_{12}, r_{3}; k_{12}, k_{3})  ,
\end{align}
where  
 \begin{align}
 \psi^{(in)}_{(\gamma)} & (r_{\alpha \beta}, r_{\gamma};q_{ij}, q_{k})     \nonumber \\
&   =  \int_{-\infty}^{\infty}   \frac{d k}{2\pi}   e^{i \sqrt{\sigma^{2}    -\frac{3}{4} k^{2}} | r_{\alpha \beta} | }e^{i k  r_{\gamma}  }      i t_{\alpha \beta}(\sqrt{\sigma^{2}   -\frac{3}{4} k^{2} } )   \nonumber \\
&     \quad    \times  \int_{-\infty}^{\infty}   d r'_{\gamma}    e^{ - i k r'_{\gamma} } \psi_{(0)}(0, r'_{\gamma};q_{ij}, q_{k})         ,
\end{align}
and
\begin{align}
& \phi^{*}_{(U)}  (r_{12}, r_{3}; k_{12}, k_{3}) =   e^{-i k_{12} r_{12}}e^{-i k_{3} r_{3}}   \nonumber \\
&\quad \quad \quad \quad + \int_{-\infty}^{\infty}  d r'_{12} d r'_{3} e^{-i k_{12} r'_{12}}e^{-i k_{3} r'_{3}}   \nonumber \\
& \quad\quad\quad   \quad \quad   \times   m U(r' )     G_{(U)} (r'_{12}, r'_{3}; r_{12}, r_{3}; \sigma)  .
\end{align}
The wave function $\phi_{(U)}  $ satisfies Schr\"odinger  equation with the presence of three-body forces potential alone,
\begin{align}
& \left [  \sigma^{2}  + \frac{d^{2}}{d r^{2}_{12}} + \frac{3}{4} \frac{d^{2}}{d r^{2}_{3}}  - m U(r )    \right ]    \phi_{(U)}  (r_{12}, r_{3}; q_{ij}, q_{k}) =0  .
\end{align}

The finite volume three-body wave function again can be constructed from three-body free space wave function, see  Eq.(\ref{relwaveconstr}),  therefore, when three-body force is considered, we obtain the finite volume three-body wave function,
 \begin{align}
 \psi^{(L)} & (r_{12}, r_{3}; q_{ij},q_{k})   = \sum_{\gamma=1}^{3} \psi^{(L)}_{(\gamma)}   (r_{\alpha \beta}, r_{\gamma};q_{ij}, q_{k}) \nonumber \\
&  -  \frac{1}{L^{2}}    \sum_{(n_{12}, n_{3})\in \mathbb{Z}}^{ \substack{ k_{3} = - \frac{2P}{3} + \frac{2\pi}{L} n_{3}, \\ k_{12} = - \frac{P}{3} - \frac{k_{3}}{2} + \frac{2\pi}{L} n_{12}} }    e^{i k_{12} r_{12}}   e^{i k_{3}  r_{3} }    \nonumber \\
& \quad \quad \quad  \quad \quad  \times    \frac{  T_{(U)} (k_{12}, k_{3}; q_{ij}, q_{k})  }{\sigma^{2} - k_{12}^{2} - \frac{3}{4} k_{3}^{2} + i \epsilon} ,
\end{align}
where $\psi^{(L)}_{(\gamma)}$ is given by Eq.(\ref{finitewaveboth}).


\begin{thebibliography}{99}







\bibitem{Kambor:1995yc}
J.~Kambor, C.~Wiesendanger, and D.~Wyler, Nucl.\ Phys.\ {\bf B465},  215  (1996).

\bibitem{Anisovich:1996tx}
A.~V.~Anisovich and H.~Leutwyler, Phys.\ Lett.\ {\bf B375},  335  (1996).

\bibitem{Colangelo:2009db}
G.~Colangelo, S.~Lanz, and E.~Passemar, PoS {\bf CD09},  047  (2009).

\bibitem{Lanz:2013ku}
S.~Lanz, PoS {\bf CD12},  007  (2013).



\bibitem{Schneider:2010hs}
S.~P.~Schneider, B.~Kubis, and C.~Ditsche, JHEP {\bf 1102},  028  (2011).

\bibitem{Kampf:2011wr}
K.~Kampf, M.~Knecht, J.~Novotny, and M.~Zdrahal, Phys.\ Rev.\ {\bf D84},  114015
  (2011).
 


\bibitem{Guo:2015zqa}
P.~Guo, Igor~V.~Danilkin, D.~Schott, C.~Fern\'andez-Ram\'{\i}rez,  V.~Mathieu and A.~P.~Szczepaniak, Phys.\ Rev.\ {\bf D92},  054016
  (2015).


\bibitem{Guo:2016wsi}
P.~Guo, Igor~V.~Danilkin, C.~Fern\'andez-Ram\'{\i}rez,  V.~Mathieu and A.~P.~Szczepaniak, Phys.\ Lett.\ {\bf B771},  497
  (2017).


 








\bibitem{Taylor:1966zza} 
 J.~G.~Taylor,
Phys.\ Rev.\  {\bf 150}, 1321 (1966).



\bibitem{Basdevant:1966zzb} 
 J.~-L.~Basdevant and R.~E.~Kreps,
Phys.\ Rev.\ {\bf 141}, 1398 (1966).



\bibitem{Gross:1982ny} 
 F.~Gross,
Phys.\ Rev.\ C {\bf 26}, 2226 (1982).








  \bibitem{Faddeev:1960su}
	L.~D.~Faddeev,
  Zh.\ Eksp.\ Teor.\ Fiz.\  {\bf 39}, 1459 (1960) [Sov. Phys.-JETP {\bf 12}, 1014(1961)]. 







  \bibitem{Faddeev:1965}
	L.~D.~Faddeev,
  {\it Mathematical Aspects of the Three-Body Problem in the Quantum Scattering Theory}, Israel Program for Scientific Translation, Jerusalem, Israel (1965).





\bibitem{Gloeckle:1983}
W.~Gl\"ockle, 
 {\it The Quantum Mechanical Few-Body Problem}, Springer, Berlin, Germany (1983).



 \bibitem{Phillips:1966zza}
A.~C.~Phillips,
Phys.\ Rev.\  {\bf 142}, 984  (1966).



 \bibitem{Fedorov:1993}
D.~V.~Fedorov and A~S.~Jensen,
Phys.\ Rev.\ Lett.\  {\bf 71}, 4103  (1993).


\bibitem{Gloeckle:1995jg}
W.~Gl\"ockle,  H.~Witala, D.~H\"uber, H.~Kamada and J.~Golak, 
Phys.\ Rept.\   {\bf 274}, 107 (1996).









\bibitem{Khuri:1960zz} 
  N.~N.~Khuri  and S.~B.~Treiman,
  Phys.\ Rev.\  {\bf 119}, 1115 (1960).


\bibitem{Bronzan:1963xn} 
  J.~B.~Bronzan   and C.~Kacser,
  Phys.\ Rev.\  {\bf 132}, 2703 (1963).



\bibitem{Aitchison:1965kt} 
  I.~J.~R.~Aitchison,
  II\ Nuovo\ Cimento\ {\bf 35}, 434 (1965).





\bibitem{Aitchison:1965zz} 
  I.~J.~R.~Aitchison,
  Phys.\ Rev.\ {\bf 137}, B1070 (1965); \ Phys.\ Rev.\ {\bf 154}, 1622 (1967).




\bibitem{Aitchison:1966kt} 
  I.~J.~R.~Aitchison and R.~Pasquier,
  Phys.\ Rev.\ {\bf 152}, 1274 (1966).



\bibitem{Pasquier:1968zz} 
 R.~Pasquier and  J.~Y.~Pasquier,
  Phys.\ Rev.\ {\bf 170}, 1294 (1968).
  


\bibitem{Pasquier:1969dt} 
 R.~Pasquier and  J.~Y.~Pasquier,
  Phys.\ Rev.\ {\bf 177}, 2482 (1969).





    
\bibitem{Guo:2014vya} 
  P.~Guo, I.~V.~Danilkin  and  A.~P.~Szczepaniak,
 Eur.\ Phys.\ J. \ A {\bf 51}, 135 (2015).





\bibitem{Guo:2014mpp} 
  P.~Guo,
  Phys.\ Rev.\ D {\bf 91}, 076012 (2015).


    
\bibitem{Danilkin:2014cra} 
  I.~V.~Danilkin, C.~Fern\'andez-Ram\'{\i}rez, P.~Guo, V.~Mathieu, D.~Schott and A.~P.~Szczepaniak,
  Phys.\ Rev.\ D {\bf 91}, 094029 (2015).




\bibitem{Guo:2015kla} 
  P.~Guo,
  Mod.\ Phys.\ Lett.\ A {\bf 31}, 1650058 (2016).
















            
            \bibitem{Aoki:2007rd} 
  S.~Aoki {\it et al.}  [CP-PACS Collaboration],
  Phys.\ Rev.\ D {\bf 76}, 094506 (2007)
  
            
  \bibitem{Sasaki:2008sv}
	K.~Sasaki, and N.~Ishizuka,
  Phys.\ Rev.\ D {\bf 78}, 014511 (2008).

  \bibitem{Feng:2010es}
	X.~Feng, K.~Jansen, and D.~B.~Renner,
  Phys.\ Rev.\ D {\bf 83}, 094505 (2011).

  \bibitem{Dudek:2010ew}
	J.~J.~Dudek  {\it et al.}  (Hadron Spectrum Collaboration),
  Phys.\ Rev.\ D {\bf 83}, 071504 (2011).

  \bibitem{Beane:2011sc}
	S.~R.~Beane  {\it et al.}  (NPLQCD Collaboration),
	Phys.\ Rev.\ D {\bf 85}, 034505 (2012).


  \bibitem{Lang:2011mn}
	C.~B.~Lang, D.~Mohler, S.~Prelovsek and M.~Vidmar,
	Phys.\ Rev.\ D {\bf 84}, 054503 (2011).

\bibitem{Aoki:2011yj} 
  S.~Aoki {\it et al.}  [CS Collaboration],
  Phys.\ Rev.\ D {\bf 84}, 094505 (2011)
  
  




  \bibitem{Dudek:2012gj}
	J.~J.~Dudek  {\it et al.}  (Hadron Spectrum Collaboration),
	Phys.\ Rev.\ D {\bf 86}, 034031 (2012).




  \bibitem{Dudek:2012xn}
	J.~J.~Dudek, R.~G.~Edwards and  C.~E.~Thomas,
	Phys.\ Rev.\ D {\bf 87}, 034505 (2013).





  \bibitem{Wilson:2014cna}
	D.~J.~Wilson,   J.~J.~Dudek, R.~G.~Edwards and  C.~E.~Thomas,
	Phys.\ Rev.\ D {\bf 91},  054008 (2015).






  \bibitem{Wilson:2015dqa}
	D.~J.~Wilson, R.~A.~Briceno,  J.~J.~Dudek, R.~G.~Edwards and  C.~E.~Thomas,
	Phys.\ Rev.\ D {\bf 92},  094502 (2015).






  \bibitem{Dudek:2016cru}
	J.~J.~Dudek,   {\it et al.}  (Hadron Spectrum Collaboration),
	Phys.\ Rev.\ D {\bf 93},  094506 (2016).







 


 



  \bibitem{Luscher:1990ux}
M.~L\"uscher,
  Nucl.\ Phys.\ B {\bf 354}, 531 (1991).



  \bibitem{Rummukainen:1995vs}
K.~Rummukainen and S.~Gottlieb,
  Nucl.\ Phys.\ B {\bf 450}, 397 (1995).

  \bibitem{Lin:2001ek}
C.-J.~D.~Lin, G.~Martinelli, C.~T.~Sachrajda and M.~Testa,
  Nucl.\ Phys.\ B {\bf 619}, 467 (2001).
            
  \bibitem{Christ:2005gi}
	N.~H.~Christ, C.~Kim  and T.~Yamazaki,
  Phys.\ Rev.\ D {\bf 72}, 114506 (2005).
            
  \bibitem{Bernard:2007cm}
V.~Bernard, Ulf-G.~Mei{\ss}ner and A.~Rusetsky,
  Nucl.\ Phys.\ B {\bf 788}, 1 (2008).
  
  \bibitem{Bernard:2008ax}
V.~Bernard, M.~Lage, Ulf-G.~Mei{\ss}ner and A.~Rusetsky,
  JHEP {\bf 0808}, 024 (2008).
  
   
  \bibitem{He:2005ey}
S.~He, X.~Feng and C.~Liu,
  JHEP {\bf 0507}, 011 (2005).




  \bibitem{Lage:2009zv}
M. Lage, Ulf-G. Mei{\ss}ner and A. Rusetsky,
Phys.\ Lett.\ B {\bf 681}, 439 (2009)




  \bibitem{Doring:2011vk}
M.~D\"oring, Ulf-G.~Mei{\ss}ner, E.~Oset and A.~Rusetsky,
Eur.\ Phys.\ J.\ A {\bf 47}, 139 (2011)




\bibitem{Aoki:2011gt} 
  S.~Aoki {\it et al.}  [HAL QCD Collaboration],
  Proc.\ Japan Acad.\ B {\bf 87}, 509 (2011)
  
    

\bibitem{Briceno:2012yi} 
  R.~A.~Briceno and Z.~Davoudi,
  Phys.\ Rev.\ D {\bf 88}, 094507 (2013)
  



\bibitem{Hansen:2012tf} 
  M.~T.~Hansen and S.~R.~Sharpe,
  Phys.\ Rev.\ D {\bf 86}, 016007 (2012)
  



\bibitem{Guo:2012hv} 
  P.~Guo, J.~Dudek, R.~Edwards  and A.~P.~Szczepaniak,
  Phys.\ Rev.\ D {\bf 88}, 014501 (2013)
  


  \bibitem{Guo:2013vsa}
  P.~Guo, 
  Phys.\ Rev.\  D {\bf 88}, 014507 (2013).









\bibitem{Rei02} S.~M.~Reimann and M.~Manninen, Rev.\ Mod.\ Phys.\  {\bf 74}, 1283 (2002).

\bibitem{Jef02} J.~H.~Jefferson, M.~Fearn, D.~L.~J.~Tipton and T.~P.~Spiller, Phys.\ Rev.\  A  {\bf 66}, 042328  (2002). 

\bibitem{Gol08} V.~N.~Golovach, A.~Khaetskii and D.~Loss, Phys.\  Rev.\  B {\bf 77},  045328  (2008).

\bibitem{Sch13} S.~Schr\"oter, P.~-A.~Hervieux, G.~Manfred, J.~Eiglsperger and J.~Madro\"nero, Phys.\  Rev.\  B  {\bf  87}, 155413 (2013).

\bibitem {Fros13} T.~Frostad, J.~P.~Hansen, C.~J.~Wessl\"en, E.~Lindroth and E.~Rasanen, Eur.\ Phys.\ J.\  B {\bf 86}, 430 (2013).

\bibitem {Sar10} J.~Sarkka and A.~Harju, Physica E {\bf 42}, 844 (2010).

\bibitem {Wang11} Z.~-W.~Wang and S.~-S.~Li, Solid State Commun.\ {\bf 151}, 1667 (2011).

\bibitem {Ahn14} J.~S.~Ahn, Superlattices Microstruct.\  {\bf 65}, 113 (2014).

\bibitem {Men11} M.~Mengesha and V.~N.~Mal{\rq}nev, Ukr.\  J.\  Phys.\ {\bf 56} (11), 1228 (2011). 


\bibitem {Khr13} R.~Khordad, Superlattices Microstruct.\  {\bf 62}, 166 (2013).

\bibitem {Ful11} M.~R.~Fulla, F.~Rodriguez-Prada, J.~H.~Mar'n~Cadavid, Superlattices Microstruct.\  {\bf 49}, 252 (2011).

\bibitem {Fros14} L.~F.~Garcia, W.~GutiŽrrez and I.~D.~Mikhailov, Physica B {\bf 455}, 14 (2014).

\bibitem {Yak15} Y.~Yakar, B.~'akir and A.~\"Ozmen, Comput.\ Phys.\ Commun.\ {\bf 188}, 88 (2015).

\bibitem {Gak13} B.~Gakir, Y.~Yakar, A.~\"Ozmen, Physica B {\bf 60}, 389 (2013).



\bibitem {Lev92} A.~Aharony, O.~Entin-Wohlman,Y.~Levinson and Y.~Imry,  Ann.\  Phys.\  (Leipzig) {\bf 8}, 685 (1999).

 
\bibitem {Alt85} B.~L.~Altshuler and A.~G.~Aronov, {\it Modern Problems in Condensed Matter Sciences}, {\bf 10},   1-153 (1985).



\bibitem {Hus14}   D.~A.~Huse, R.~Nandkishore and V.~Oganesyan,  Phys.\  Rev.\  B  {\bf  90}, 174202  (2014).





  \bibitem{Kreuzer:2008bi}
	S.~Kreuzer and H.-W.~Hammer,
	Phys.\ Lett.\ B {\bf 673},  260 (2009).
	
	

  \bibitem{Kreuzer:2009jp}
	S.~Kreuzer and H.-W.~Hammer,
	Eur.\ Phys.\ J.\ A {\bf 43},  229 (2010).
	
	

  \bibitem{Kreuzer:2012sr}
	S.~Kreuzer and H.-W.~Hammer,
	Eur.\ Phys.\ J.\ A {\bf 48},  93 (2012).
		


  \bibitem{Polejaeva:2012ut}
	K.~Polejaeva and A.~Rusetsky,
	Eur.\ Phys.\ J.\ A {\bf 48},  67 (2012).
		



  \bibitem{Briceno:2012rv}
	 R.~A.~Briceno and  Z.~Davoudi,
	Phys.\ Rev.\ D {\bf 87},  094507 (2013).





  \bibitem{Hansen:2014eka}
	 M.~T.~Hansen and S.~R.~Sharpe,
	Phys.\ Rev.\ D {\bf 90},  116003 (2014).





  \bibitem{Hansen:2015zga}
	 M.~T.~Hansen and S.~R.~Sharpe,
	Phys.\ Rev.\ D {\bf 92},  114509 (2015).



  \bibitem{Hansen:2016fzj}
	 M.~T.~Hansen and S.~R.~Sharpe,
	Phys.\ Rev.\ D {\bf 93},  096006 (2016).






\bibitem{Hammer:2017uqm}
H.~-W.~Hammer, J.~-Y.~Pang and A.~Rusetsky, arXiv:1706.07700 [hep-lat] .



\bibitem{Hammer:2017kms}
H.~-W.~Hammer, J.~-Y.~Pang and A.~Rusetsky, arXiv:1707.02176 [hep-lat] .



\bibitem{Guo:2016fgl}
P.~Guo, Phys.Rev. {\bf D95},  054508 (2017).




\bibitem{Guo:2017ism}
P.~Guo and V.~Gasparian, arXiv:1701.00438 [hep-lat] .


















  







 
 

  \bibitem{McGuire:1964zt}
J.~B.~McGuire,
  J.\ Math.\ Phys.\  {\bf 5}, 622 (1964).




  \bibitem{McGuire:1988}
  J.~B.~McGuire and C.~A.~Hurst,
  J.\ Math.\ Phys.\  {\bf 29}, 155 (1988).


  
  





 





 




 










  \bibitem{Dodd:1970}
  L.~R.~Dodd, 
  J.\ Math.\ Phys.\    {\bf 11}, 207 (1970).




 
  \bibitem{Majumdar:1972}
  C.~K.~Majumdar, 
  J.\ Math.\ Phys.\    {\bf 13}, 705 (1972).




 





  
    
\end{thebibliography}
\end{document}